\documentclass[10pt,a4paper,english,pra, twocolumn]{revtex4-1}
\usepackage[T1]{fontenc}
\usepackage[latin9]{inputenc}
\usepackage{color}
\definecolor{page_backgroundcolor}{rgb}{1, 1, 1}
\pagecolor{page_backgroundcolor}
\usepackage{float}
\usepackage{amsmath}
\usepackage{amssymb}
\usepackage{graphicx}

\makeatletter

\usepackage[colorlinks,
            linkcolor=blue,
            anchorcolor=blue,
            citecolor=blue
            ]{hyperref}

\makeatother

\usepackage{babel}
\begin{document}
\title{Non-Gaussian state preparation and enhancement using weak-value amplification }
\author{Xiao-Xi Yao }
\author{Yusuf Turek}
\email{Corresponding author: yusufu1984@hotmail.com}

\affiliation{$^{1}$School of Physics,Liaoning University,Shenyang,Liaoning 110036,China}

\maketitle
\textbf{ABSTRACT }

We introduce a protocol for generating non-Gaussian (nG) states via
postselected weak measurement. The scheme involves injecting an arbitrary
quantum state and a single photon into the signal and idler ports
of an interferometer with a third-order nonlinear medium. An nG state
is conditionally produced at the signal output, heralded by single-photon
detection in an idler output channel. The protocol exploits a weak
cross-Kerr interaction, with effective single-photon nonlinearity
enhanced by weak-value amplification. By tuning the weak value of
the idler photon number operator within experimentally feasible parameters,
diverse nG states can be generated with high fidelity. Specific examples
include photon-added coherent states, displaced and squeezed number
states, and intermediate nG states from coherent and squeezed vacuum
inputs. Furthermore, the protocol enables enhancement of non-Gaussianity
and enlargement of Schr\"{o}dinger cat (SC) states when ideal SC
states are used as input. Our results provide an alternative route
for conditional generation of tunable nG states, with potential applications
in quantum information processing and state engineering.

\textbf{INTRODUCTION }

Non-Gaussian (nG) states \citep{PRXQuantum.2.030204,PRXQuantum.1.020305},
defined as quantum states in the phase space that deviate from a Gaussian
distribution, are often characterized by Wigner function negativity
and have garnered significant interest due to their enhanced utility
across various domains of quantum physics. In particular, they play
a central role in continuous-variable (CV) quantum information processing
\citep{RevModPhys.77.513}, including quantum computation \citep{PhysRevLett.82.1784,PhysRevA.64.012310,PhysRevLett.97.110501,PhysRevA.93.022301,PhysRevLett.109.230503},
quantum metrology \citep{PhysRevLett.104.103602,PhysRevA.78.063828},
and foundational studies in quantum theory \citep{PhysRevA.67.012106,PhysRevA.75.052105}.
The generation of nG states requires either nG initial states or nG
measurements; the latter alone is sufficient to produce non-Gaussianity.
Among the existing techniques, nG unitary transformations and conditional
measurements are the two most common approaches \citep{PRXQuantum.2.030204}.
Over the past decades, considerable efforts have focused on preparing
nG states using conditional measurements involving nG quantum operations,
such as photon addition and/or subtraction applied to Gaussian states
in optical systems \citep{PhysRevA.82.063833,2004,Wakui:07,PhysRevLett.97.083604,PhysRevLett.92.153601,PhysRevA.110.023703}. 

Squeezed single-photon and single-photon-added coherent (SPAC) states
are among the most representative nG states and play irreplaceable
roles in CV quantum information processing. These states can be prepared
via nG operations such as single-photon addition and subtraction,
applied to coherent and squeezed vacuum (SV) states, respectively.
Another paradigmatic example is the Schr\"{o}dinger cat (SC) state
\citep{2006} inspired by Schr\"{o}dinger's famous thought experiment
\citep{schrodinger1935gegenwartige}. Current methods for generating
SC states include photon-number-resolving detection \citep{PhysRevA.55.3184,PhysRevA.59.1658}
and weak and strong nonlinear optical interactions \citep{PhysRevLett.57.13,PhysRevA.59.4095}.
Large-amplitude SC states characterized by coherent amplitudes $\vert\alpha\vert\ge2$,
where $\alpha$ is the complex amplitude, are particularly valuable
for probing quantum foundations and enabling quantum information protocols
\citep{PhysRevA.64.022313,PhysRevA.68.022321,PhysRevA.69.022315}.
However, SC states generated via photon subtraction or third-order
optical nonlinearities typically have small amplitudes and fall short
of the requirements imposed by advanced quantum tasks.

From a practical perspective, the purity of a state is critical for
effective implementation of many quantum information protocols. However,
due to limitations in photon detection efficiency and photon-number-resolving
capabilities, generating high-fidelity nG states remains significantly
more challenging than producing their Gaussian counterparts. This
difficulty underscores the limitations of existing methods for preparing
high-quality nG states. To address this challenge, one can either
enhance the non-Gaussianity of a given state through optimization
techniques or develop alternative approaches capable of generating
nG states with high purity and fidelity tailored for specific applications. 

In this study, we propose an alternative protocol for generating a
wide range of nG states based on the technique of weak-value amplification
(WVA) \citep{PhysRevLett.60.1351,RevModPhys.86.307}, without the
need for explicit photon addition or subtraction operations. Our scheme
relies on a weak cross-Kerr interaction between two optical modes,
the signal and idler, mediated by third-order nonlinear susceptibility
$\chi^{(3)}$. A nG state is conditionally generated in the signal
output port from a given input signal state, heralded by single-photon
detection in the idler output mode. The key element of the protocol
is a postselected weak measurement (WM) of the photon number of the
idler beam, which effectively induces single-photon nonlinearity via
the WVA. 

Although the protocol features a low probability of success, it enables
the high-fidelity generation of various nG states. By appropriately
tuning the weak value of the photon number operator in the idler mode
within experimentally feasible parameters, our protocol can generate
photon-added states and displaced number states from coherent input
states. In the case of a SV input, the scheme yields high-fidelity
two-photon-added squeezed vacuum (TPASV) states, squeezed number states,
and equivalent squeezed even SC states. When applied to SC input states,
the protocol serves two distinct purposes: (i) it significantly enhances
the non-Gaussianity of the initial state, especially for small-amplitude
even SC states, and (ii) it enables the expansion of small SC states
into large-amplitude ($|\alpha|\geq2$) SC states while preserving
high fidelity ($F>0.99$). Additionally, our approach supports the
generation of a continuous family of nonclassical states with intermediate
properties corresponding to different classes of input states. 

\vspace{0.3cm}
\textbf{RESULTS}

\textbf{A Versatile Protocol via Weak Measurement}

A previous study \citep{PhysRevA.105.022608} introduced a quantum
state engineering scheme based on postselected WMs implemented in
a medium characterized by the second-order nonlinear susceptibility
of a BBO crystal. The associated interaction Hamiltonian is given
by $H_{I}=\xi(a^{\dagger}b^{\dagger}-ab)$, where $a$ ($a^{\dagger}$)
and $b$ ($b^{\dagger}$) are the annihilation (creation) operators
of the signal and idler modes, respectively. In that proposal, the
interaction strength $\chi^{(2)}$ is proportional to the second-order
nonlinear susceptibility of the medium, that is $\xi\propto\chi^{(2)}$,
and the signal and idler modes play the roles of the pointer and the
measured system, respectively. In contrast, in many quantum optical
protocols involving light-matter interactions, a different von Neumann-type
interaction Hamiltonian is often employed, namely, the cross-Kerr
interaction, described by 
\begin{equation}
H_{int}=\chi^{(3)}n_{a}\otimes n_{b},\label{eq:17}
\end{equation}
where $n_{a}=a^{\dagger}a$ and $n_{b}=b^{\dagger}b$ are the number
operators for the signal and idler modes, respectively, and $\chi^{(3)}$
denotes the third-order nonlinear susceptibility of the medium. As
demonstrated in previous works \citep{PhysRevLett.107.133603,Nature2017,RN11},
if the signal and idler beams are identified as the measured system
and the pointer, respectively, the interaction Hamiltonian $H_{int}$
can be used within a von Neumann-type measurement framework. Analogously
to the standard WM scenario, in the Hamiltonian $H_{int}$, the photon
number operator $n_{a}$ serves as the observable of the measured
system, while $n_{b}$ acts as the pointer variable, canonically conjugate
to the phase operator $\varphi$ of the radiation field. In this section,
we demonstrate the utility of the interaction Hamiltonian $H_{int}$
for preparation of quantum states via a postselected WM technique.
\begin{figure}
\includegraphics[width=8cm]{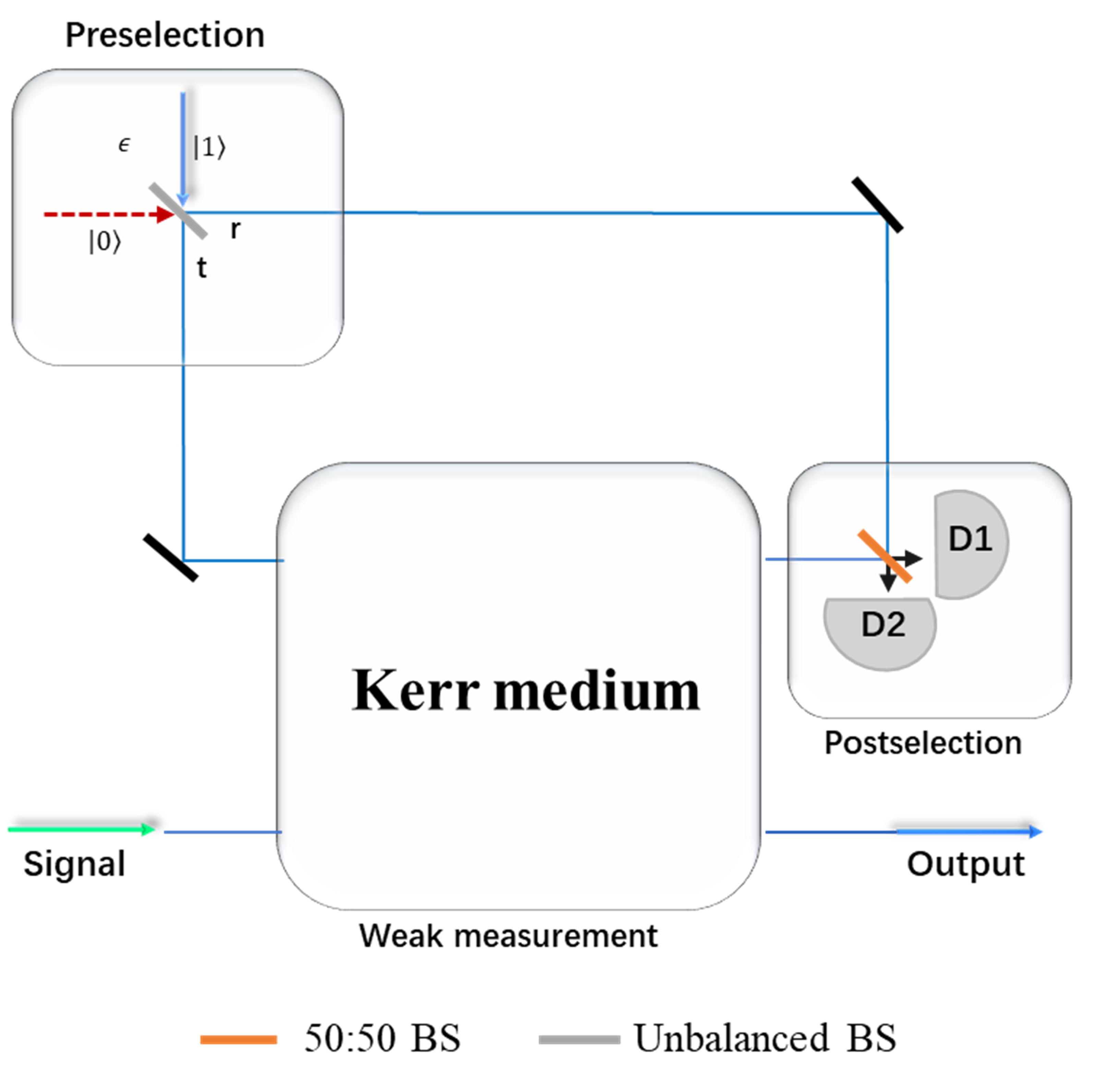}

\caption{\label{fig:5} \textbf{Schematic of the nG state generation protocol}.
In this scheme, a given state is prepared in the input signal mode,
and a single-photon state is prepared in the idler mode, both of which
pass through a MZI. Preselection is performed by sending the single
photon through an unbalanced BS with a small deviation $\epsilon$
from the ideal angle $\pi/4$. The signal and idler modes are then
coupled via a Kerr medium. A single-photon detection event at one
of the idler output ports (i.e., a click at detector $D_{1}$) heralds
the desired output state in the output signal mode. }
\end{figure}

The schematic setup of our state generation model, based on a postselected
WM characterized by the Hamiltonian interaction $H_{int}$, is shown
in Fig. \ref{fig:5}. As illustrated, the proposed setup consists
of a Mach-Zehnder interferometer (MZI), Kerr medium, and several optical
elements. In our model, the beam splitter (BS) plays a crucial role
in the execution of the protocol. The action of a two-mode BS is described
by the scattering matrix $U_{BS}$ \citep{PhysRevA.40.1371,Agarwal2013}
with 
\begin{equation}
U_{BS}=\left(\begin{array}{cc}
\cos\theta e^{i\varphi_{\tau}}\ \  & \sin\theta e^{i\phi_{\rho}}\ \\
-\sin\theta e^{-i\phi_{\rho}} & \cos\theta e^{-i\varphi_{\tau}}
\end{array}\right),\label{eq:16}
\end{equation}
where the reflectance and transmittance of the BS are given by $\sqrt{R}=\sin\theta$
and $\sqrt{T}=\cos\theta$, respectively, satisfying $R+T=1$. We
assume that the signal mode is initially prepared in an arbitrary
quantum state $\vert\phi\rangle$ and passes through the Kerr medium,
while the idler mode is initialized in a single-photon state. As shown
in Fig. \ref{fig:5}, when the input to the first BS is $\vert1\rangle_{1}\vert0\rangle_{2}$,
the resulting state after the BS is 
\begin{align}
\vert\psi_{i}\rangle & =\cos\theta_{1}\vert0\rangle_{r}\vert1\rangle_{t}+i\sin\theta_{1}\vert1\rangle_{r}\vert0\rangle_{t},\label{eq:20-1}
\end{align}
assuming $\varphi_{\tau}=0$, and $\phi_{\rho}=\pi/2$. The subindices
$r$ and $t$ of the states represent the existence of single-photon
in reflected and transmitted paths of the interferometer after the
first BS. The interaction between the measured system and pointer
occurs via the Kerr medium, described by the time-evolution operator:
\begin{align}
U & =\exp\left(-\frac{i}{\hbar}\int_{0}^{t}H_{I}d\tau\right)=\exp\left(-ign_{a}\otimes n_{b}\right),\label{eq:21}
\end{align}
where $g=\chi^{(3)}t$ is the interaction strength, with $t=L/\upsilon$
being the interaction time, where $L$ is the length of the Kerr medium,
and $\upsilon$ is the speed of light in the medium. If $g$ is typically
small and mean photon numbers in $a$ and $b$ modes both not large,
it suffices to expand the evolution operator to the first order as
follows: 
\begin{equation}
U=\exp\left(-ign_{a}\otimes n_{b}\right)\approx1-ign_{a}\otimes n_{b}.\label{eq:22}
\end{equation}
Under this approximation, the initial state of the composite system
evolves as 
\begin{align}
\vert\Psi^{\prime}\rangle & =U\vert\psi_{i}\rangle\otimes\vert\phi\rangle=\exp\left(-ign_{a}\otimes n_{b}\right)\vert\psi_{i}\rangle\otimes\vert\phi\rangle\nonumber \\
 & \approx\left[1-ign_{a}\otimes n_{b}\right]\vert\psi_{i}\rangle\otimes\vert\phi\rangle.\label{eq:No=00003D000020post}
\end{align}
Since the idler mode in our scheme contains only one photon, the validity
of this approximation requires $g\langle n_{b}\rangle\ll1$ and $\frac{1}{2}g^{2}\langle n_{b}^{2}\rangle\ll1$.
These conditions can be satisfied with weak nonlinearities and input
states of moderate photon numbers. Here, $\langle n_{b}\rangle$ and
$\langle n_{b}^{2}\rangle$ represent the expectation values of the
operators $n_{b}$ and $n_{b}^{2}$ for the initial signal state $\vert\phi\rangle$,
respectively. 

We now proceed to the postselection stage. After passing through a
balanced ($50:50$) BS, we require that detector $D1$ registers a
single photon while detector $D2$ remains inactive. This detection
event corresponds to projecting onto the state:
\begin{align}
|\psi_{f}\rangle & =\text{\ensuremath{\frac{1}{\sqrt{2}}(|0\rangle_{r}|1\rangle_{t}-i|1\rangle_{r}|0\rangle_{t})},}\label{eq:24}
\end{align}
where we assume $\varphi_{\tau}=0$, $\phi_{\rho}=\pi/2$ and $\theta_{2}=\pi/4$
for a balanced BS. The details of deriviations of pre- and pre-selected
states are provided in Methods. Postselecting $\vert\psi_{f}\rangle$
from the state $\vert\Psi^{\prime}\rangle$, yields the (unnormalized)
final pointer state: 
\begin{align}
\vert\Phi\rangle & =\langle\psi_{f}\vert\psi_{i}\rangle\left[1-ig\langle n_{a}\rangle_{w}n_{b}\right]\vert\phi\rangle\nonumber \\
 & =\langle\psi_{f}\vert\psi_{i}\rangle\left[\vert\phi\rangle-i\frac{g}{2}\left(1+\frac{1}{\epsilon}\right)b^{\dagger}b\vert\phi\rangle\right],\label{eq:19}
\end{align}
where 
\begin{align}
\langle n_{a}\rangle_{w} & =\frac{\langle\psi_{f}\vert a^{\dagger}a\vert\psi_{i}\rangle}{\langle\psi_{f}\vert\psi_{i}\rangle}=\frac{1}{2}+\frac{1}{2\epsilon}\label{eq:21-1}
\end{align}
is the weak value of the photon number operator $n_{a}=a^{\dagger}a$
for the idler beam. This expression assumes a slight deviation from
a $50:50$ BS, with $\theta_{1}=\pi/4-\epsilon$, where $\epsilon\ll1$
is real. For a sufficiently small $\epsilon$, the weak value can
exceed unity, as observed in previous studies \citep{PhysRevLett.107.133603,Nature2017}.
The success probability of the postselection (i.e., a click at $D1$
and no click at $D2$ ) is given by $\vert\langle\psi_{f}\vert\psi_{i}\rangle\vert^{2}=\epsilon^{2}$.
This indicates that whenever the postselection succeeds our protocol
amplifies the mean photon number of the idler mode as if there exist
more than single photon inside the interferometer. Consequently, the
signal beam experience a cross-phase shift equivalent to that of many
photons.

In the above postselection process we assume that the second BS is
ideally $50:50$ balanced. However, in practical implementation processes
of our scheme it is necessary to take into account the imperfections
of related optical elements. If both BSs are slightly unbalanced---for
example, with their transmission-to-reflection ratios characterized
by the parameters $\theta_{1}=\pi/4-\epsilon_{1}^{\prime}$ and $\theta_{2}=\pi/4\pm\epsilon_{2}^{\prime},$
respectively--the weak value obtained in above case changed to $1/2+1/2(\epsilon_{1}^{\prime}\mp\epsilon_{2}^{\prime})$,
where $\epsilon_{1}^{\prime},\epsilon_{2}^{\prime}\ll1$. Even under
such imbalanced conditions, by tuning the system parameters to satisfy
the phase-matching condition $\epsilon_{1}^{\prime}\mp\epsilon_{2}^{\prime}=\epsilon$,
one can still obtain output results identical to those of an ideal
50 : 50 BS. This demonstrates the robustness of our protocol to such
imperfections, as they can be compensated for by careful calibration
of the optical apparatus.

After the above postselected WM procedure, the normalized output state
of the signal mode is 
\begin{equation}
\vert\Phi_{f}\rangle=\frac{\vert\Phi\rangle}{\sqrt{p_{f}}},
\end{equation}
where the overall success probability $p_{f}$ of the protocol for
generating the signal output state with single-trial is given by the
general form $p_{f}=\epsilon^{2}\vert\mathcal{N}\vert^{-2}.$ Here,
$\epsilon$ is the small deviation angle of the first BS from $\pi/4$,
and $\mathcal{N}$ is the normalization constant of the final state
$\vert\Phi\rangle$, which depends on the specific input state $\vert\phi\rangle$
and system parameters ($g,\epsilon$). A key strength of our approach
is that the success probability across all types of input states follows
a universal form, allowing for a unified feasibility analysis.

From the form of $\vert\Phi_{f}\rangle$, it is evident that the output
state is a superposition of the initial state $\vert\phi\rangle$
and the state $b^{\dagger}b\vert\phi\rangle$. The interference between
these components enhances the quality of the generated state and can
lead to nontrivial quantum effects. In particular, the term $b^{\dagger}b\vert\phi\rangle$
corresponds to a photon subtraction followed by photon addition, a
process previously identified as a powerful tool for quantum state
engineering \citep{2007PV}. Therefore, when the coefficient $\frac{g}{2}\left(1+\frac{1}{\epsilon}\right)\gg1$,
the $b^{\dagger}b\vert\phi\rangle$ component dominates the output,
effectively transforming the initial state $\vert\phi\rangle$ into
a distinct quantum state with nontrivial properties and nG characteristics. 

As investigated in a recent WVA experiment \citep{PhysRevA.97.033851},
the minimal rotation angle $\epsilon$ of the BS can be as small as
$0.52$ mrad ($5.2\times10^{-4}$ rad). In a more recent study \citep{PhysRevA.111.042425},
the BS was also considered to have deviated slightly from the ideal
configuration $\pi/4$, with the deviation characterized by $\epsilon\propto10^{-3}$.
Generally, the non-linearity coefficient of the cross-Kerr $g$ is
small \citep{DONG20085677}. However, recent studies have proposed
methods to engineer giant cross-Kerr nonlinearities \citep{PhysRevLett.87.073601,2016,2018,vanDoai:19,Doai_2019,PhysRevA.103.043709,2024},
with a reported maximum value of $g=0.35$ \citep{PhysRevLett.111.053601}.
Reference \citep{Du2024kerreffectbased} explored the implementation
of various quantum logic gates based on enhanced cross-Kerr nonlinearity.
These findings suggest that the weak value $\langle n_{a}\rangle_{w}$
can be significantly amplified by WVA \citep{PhysRevLett.107.133603,Nature2017},
making it feasible to achieve the condition $\frac{g}{2}\left(1+\frac{1}{\epsilon}\right)\gg1$
with appropriate control techniques. As described, in our protocol,
the average photon number $\langle n_{b}\rangle$ is determined by
the input state $\vert\phi\rangle$. The parameters of this input
state can be adjusted to enhance the weight of the $n_{b}$ term in
the post-measurement state $\vert\Phi_{f}\rangle$ via WVA, while
maintaining the weak coupling condition $g\langle n_{b}\rangle\ll1$.
Our robust parameter analysis thereby confirms the preparation of
a new family of quantum states with our proposed scheme is experimentally
feasible.

Below, we present examples that illustrate the preparation of nG states
and the enhancement of non-Gaussianity for different input signal
states $\vert\phi\rangle$.\vspace{0.3cm}

\textbf{Non-Gaussian States Generation}

As a first example, we demonstrate the preparation of a state closely
resembling a photon-added coherent state. According to Eq.~(\ref{eq:19}),
if the input signal state is taken to be a coherent state, i.e., $\vert\phi\rangle=\vert\beta\rangle=D\left(\beta\right)\vert0\rangle$,
where the displacement operator is defined as $D(\beta)=\exp\left(\beta b^{\dagger}-\beta^{\ast}b\right)$,
then the normalized final state of the signal beam, after the postselected
WM implemented via our protocol, becomes 
\begin{align}
\vert\Phi^{\prime}\rangle & =\mathcal{N}_{c}\left[\vert\beta\rangle-ig\langle n_{a}\rangle_{w}b^{\dagger}b\vert\beta\rangle\right]=\mathcal{N}_{c}\left[\vert\beta\rangle-\kappa\vert\phi_{1}\rangle\right],\label{eq:20}
\end{align}
where $\kappa=i\beta g\langle n_{a}\rangle_{w}$, $\vert\phi_{1}\rangle=b^{\dagger}\vert\beta\rangle$,
and the normalization constant $\mathcal{N}_{c}$ is given by 
\begin{align}
\mathcal{N}_{c} & =\left[1+\vert\kappa\vert^{2}(1+\vert\beta\vert^{2})-2Re[\kappa^{\ast}\beta]\right]^{-\frac{1}{2}}.\label{eq:28}
\end{align}
Here, $Re[\mathcal{C}]$ denotes the real part of the complex number
$\mathcal{C}$. Furthermore, using the commutation relation $[b^{\dagger},D(\beta)]=\beta^{\ast}D(\beta)$,
the output signal state $\vert\Phi^{\prime}\rangle$ can be rewritten
as 
\begin{equation}
\vert\Phi^{\prime}\rangle=\mathcal{N}_{c}\left[\left(1-\kappa\beta^{\ast}\right)\vert\beta\rangle-\kappa\vert\beta,1\rangle\right],\label{eq:14}
\end{equation}
where $\vert\beta,1\rangle=D(\beta)\vert1\rangle$ denotes the displaced
single-photon state. This expression shows that state $\vert\Phi^{\prime}\rangle$
lies in the subspace spanned by two orthogonal states: the coherent
state and the displaced single-photon state. This orthogonality may
be useful for encoding CV qubits. 

\begin{figure}[H]
\includegraphics[width=8cm]{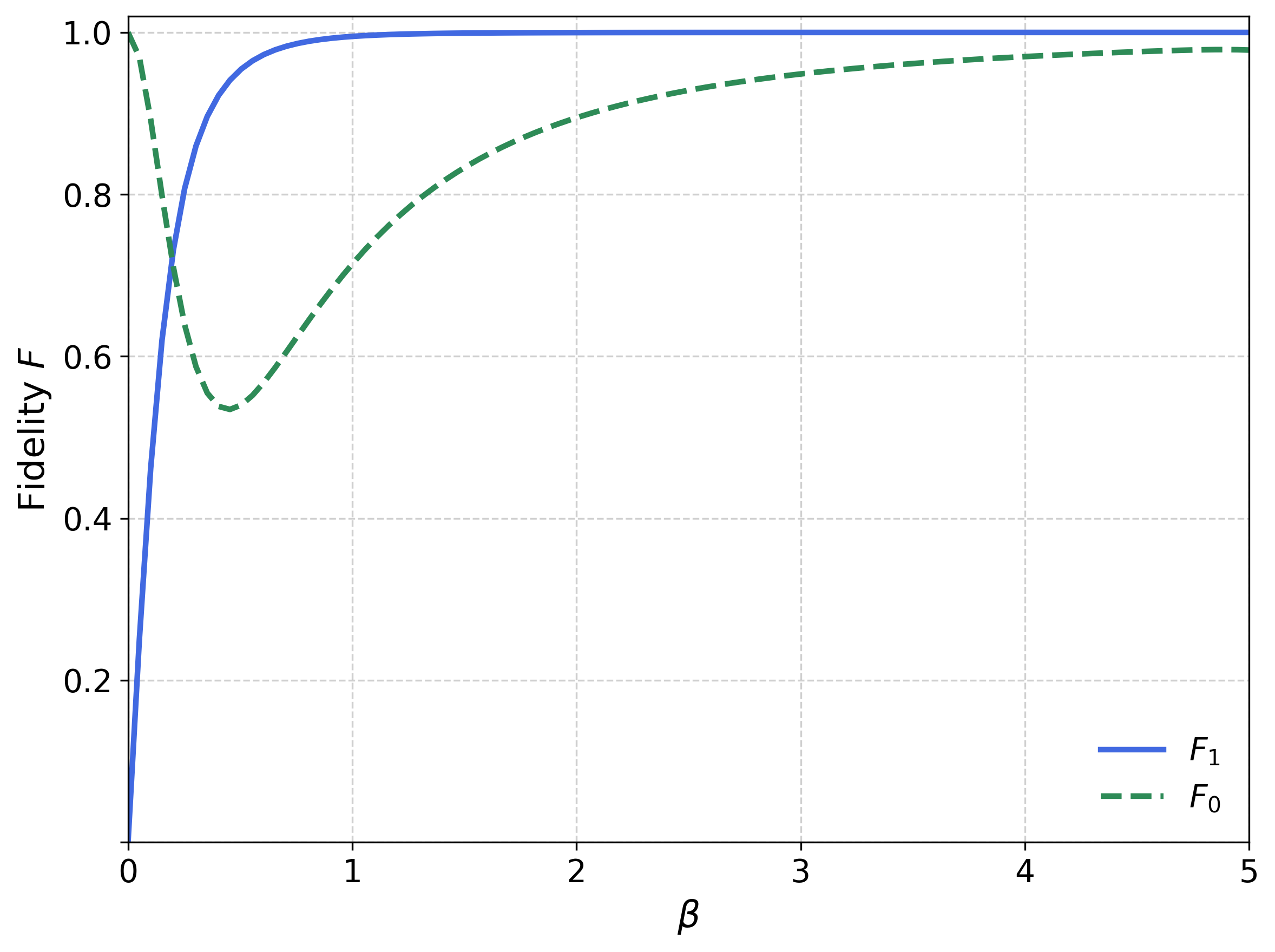}

\caption{\label{fig:2} \textbf{Fidelity between the generated output state
and reference states.} Fidelity functions between the signal output
state $\vert\Phi^{\prime}\rangle$ and the ideal SPAC state (solid
curve), as well as the coherent state $\vert\phi\rangle$ (dashed
curve). Here, $\epsilon=0.001$ and $g=0.01$.}
\end{figure}

\begin{flushleft}
We observe that the state $\vert\Phi^{\prime}\rangle$ is a superposition
of a coherent state $\vert\beta\rangle$ and a SPAC state $\vert\phi_{1}\rangle=b^{\dagger}\vert\beta\rangle$.
When the weak value $\langle n_{a}\rangle_{w}$ is large, the contribution
of the SPAC component $\vert\phi_{1}\rangle$ dominates over the coherent
component. In general, the generated state can be regarded as a typical
example of a hybrid coherent states of the form $\sqrt{\varepsilon}\vert\beta\rangle+\sqrt{1-\varepsilon}\vert\phi_{1}\rangle$,
with $0\le\varepsilon\le1$ \citep{Turek_2023}. Such states exhibit
enhanced nonclassical features compared to the SPAC state alone due
to the admixture of a coherent-state in specific parameter regimes. 
\par\end{flushleft}

\begin{flushleft}
The conventional method for generating SPAC states, as demonstrated
by Zavatta \textit{et al}. \citep{2004}, employs conditional parametric
down-conversion in a $\chi^{(2)}$ nonlinear crystal, where the detection
of an idler photon heralds the addition of a photon to a coherent
state in the signal mode. While effective, this approach is inherently
specialized for producing SPAC states and is constrained by weak second-order
nonlinearity and finite detection efficiency, resulting in a low success
probability. In contrast, our WVA based protocol, utilizing a weak
$\chi^{(3)}$ cross-Kerr interaction, offers a versatile and tunable
alternative. Although our scheme also features a low probability due
to the required postselection, it is not limited to a single state
type. By adjusting system parameters ($\beta$, $\epsilon$, g), our
single experimental setup can be tuned to prepare not only SPAC states
but also hybrid coherent states \citep{Turek_2023} and other nG states,
depending on the initial signal input.
\par\end{flushleft}

To quantify how closely the generated state $\vert\Phi^{\prime}\rangle$
resembles either the coherent state $\vert\beta\rangle$ or the ideal
SPAC state $\vert\phi_{1}\rangle$, we computed the fidelity: 
\begin{equation}
F_{0}=\vert\langle\beta\vert\Phi^{\prime}\rangle\vert^{2},\ \ F_{1}=\vert\langle\phi_{1}\vert\Phi^{\prime}\rangle\vert^{2}.\label{eq:14-1}
\end{equation}
Here, $F_{0}$ and $F_{1}$ measure the overlap between the coherent
and SPAC states, respectively. A fidelity value that is close to unity
indicates that the generated state closely approximates the corresponding
target state.

In Fig.~\ref{fig:2}, we plot $F_{0}$ and $F_{1}$ as functions
of the coherent state amplitude $\beta$. Here, we fixed the parameters
as $\epsilon=0.001$ and $g=0.01$. As seen, for small values of $\beta$,
the state is predominantly coherent, with $F_{0}$ near unity. As
$\beta$ increases, $F_{1}$ increases, indicating a closer resemblance
to the SPAC state. For a sufficiently large $\beta$, the output state
again approaches a coherent state regardless of the other parameter
values. 

To further characterize the output state, we examine its Wigner function,
which provides a complete description of the phase-space distribution
of the state. For a quantum state with density operator $\rho$, the
Wigner function is obtained via the Fourier transform of the characteristic
function $C_{w}(\lambda)=Tr\left[\rho D(\lambda)\right]$. In this
study, we used the Python package QuTiP \citep{JOHANSSON20121760,JOHANSSON20131234}
to numerically compute the Wigner functions of the relevant quantum
states.

\begin{figure}
\includegraphics[width=8cm]{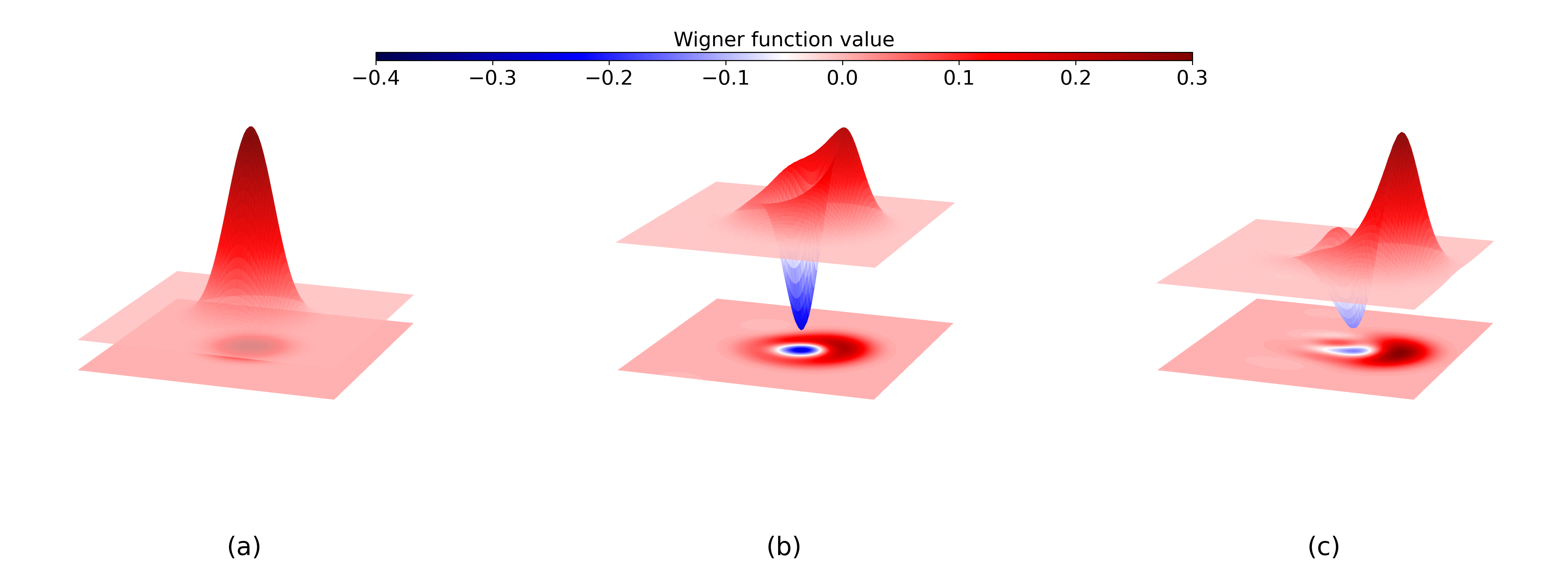}

\caption{\textcolor{black}{\label{fig:3-1}} \textbf{Wigner function of the
output state of the generated signal} $\vert\Phi^{\prime}\rangle$.
\textbf{a} $\beta=0$; \textbf{b} $\beta=0.5$; \textbf{c} $\beta=1$.
Other parameters are the same as those used in Fig.~\ref{fig:2}.}
\end{figure}

To visualize the evolution of the phase-space distribution, Fig. \ref{fig:3-1}
shows the Wigner functions of the output state $\vert\Phi^{\prime}\rangle$
for different values of $\beta$. We fixed the parameters as $\epsilon=0.001$
and $g=0.01$, and considered $\beta=0$, $0.5$, and $1$. Figure
\ref{fig:3-1}\textbf{a} corresponds to the input coherent state,
exhibiting the expected Gaussian distribution displaced in the phase
space. In Fig. \ref{fig:3-1}\textbf{b}, for $\beta=0.5$, the Wigner
function of the output state shows significant negativity near the
origin, along with squeezing in the $x$-quadrature. This indicates
strong nonclassicality and a marked deviation from the initial Gaussian
character. In Fig. \ref{fig:3-1}\textbf{c}, as $\beta$ increases
further, the negativity gradually diminishes, and the Wigner function
approaches a Gaussian profile again (not shown). This trend is consistent
with the behavior of the fidelity functions in the large-$\beta$
regime {[}see Fig.~\ref{fig:2}{]}. These numerical results demonstrate
that our protocol effectively transforms the initially Gaussian coherent
state into a nG state with tunable nonclassicality, depending on the
input amplitude $\beta$.
\begin{figure}[H]
\includegraphics[width=8cm]{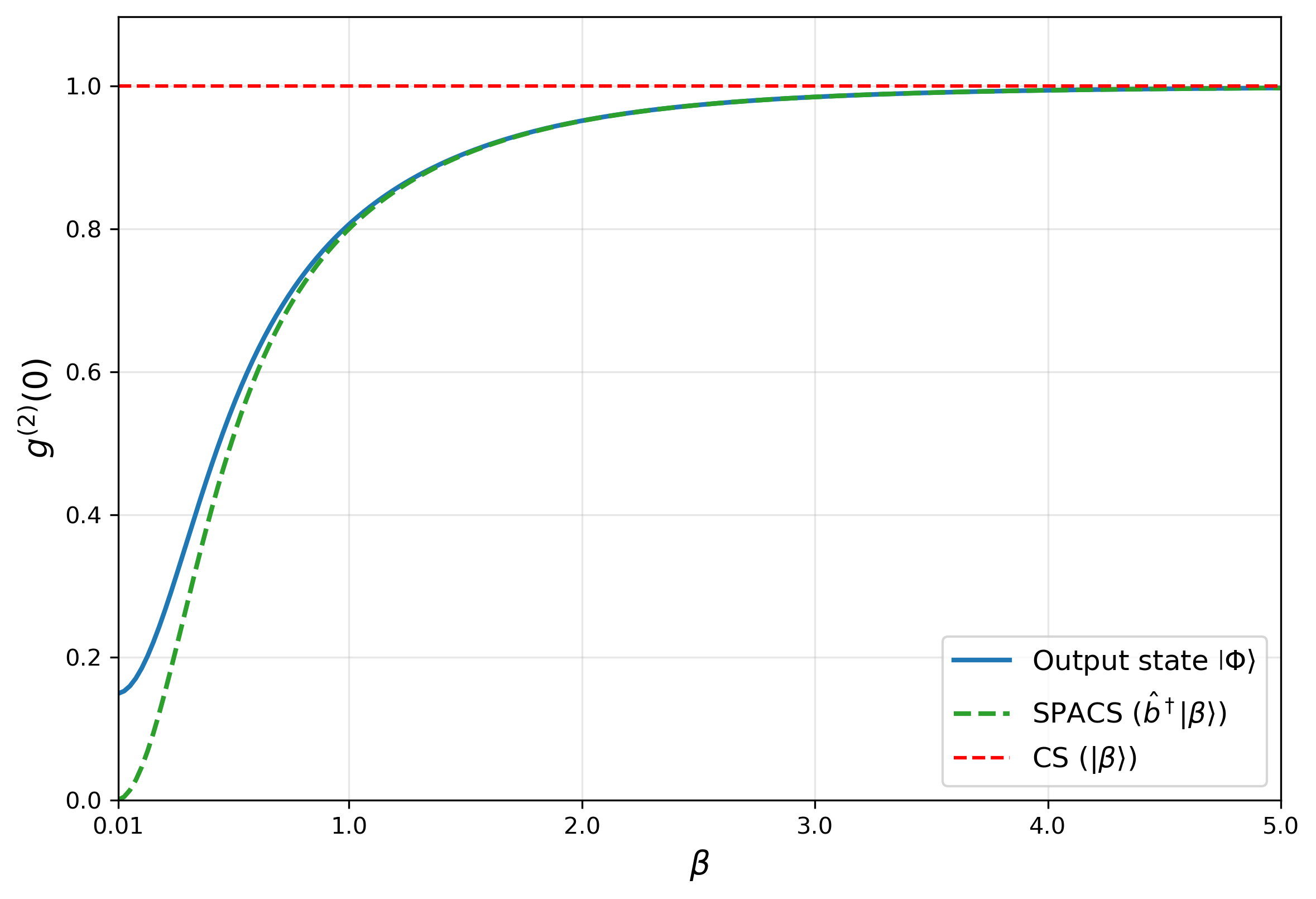}

\caption{\label{fig:4-1} \textbf{Second-order correlation function $g^{(2)}(0)$}.
The blue solid curve corresponds to the generated signal output state
$\vert\Phi^{\prime}\rangle$, while the red and green dashed curves
represent the coherent state and ideal SPAC state, respectively. Other
parameters are the same as those used in Fig.~\ref{fig:2}.}
\end{figure}

Furthermore, as shown in Fig.~\ref{fig:4-1}, the generated state
$\vert\Phi^{\prime}\rangle$ exhibits a higher single-photon component
than a coherent state, although this enhancement is not as strong
as that of the ideal SPAC state. Additionally, the state demonstrates
a pronounced sub-Poissonian photon-number distribution, as evidenced
by its photon-number variance remaining smaller than the square root
of the mean photon number, i.e., $\triangle n<\sqrt{\langle n\rangle}$. 

Since the SPAC state lacks a vacuum component and exhibits a high
probability of single-photon detection, it has found various applications
in quantum information processing, including quantum key distribution
(QKD) \citep{PhysRevA.90.062315}. In practical QKD implementations,
sources with sub-Poissonian photon-number statistics and a dominant
single-photon component are highly desirable. Based on the above analysis,
our generated state $\vert\Phi^{\prime}\rangle$ satisfies both of
these criteria, making it a promising candidate for QKD applications.
Therefore, $\vert\Phi^{\prime}\rangle$ could offer improved performance
compared to other commonly used sources, such as weak coherent states
\citep{PhysRevLett.96.070502,PhysRevLett.98.010503} and heralded
single-photon sources \citep{PhysRevLett.100.090501,PhysRevA.75.012312}.

The success probability of our signal output state $\vert\Phi^{\prime}\rangle$
is equal to $\epsilon^{2}\vert\mathcal{N}_{c}\vert^{-2}$. By numerically
analyzing the successful postselection probability of our state, we
notice that it is on the order of $10^{-5}$ for single-trial with
high fidelity ($F_{1}>0.995$). However, this low yield characteristic
is shared by many advanced quantum state generation schemes, including
the pioneering work of Zavatta \textit{et al}. \citep{2004} and Ourjoumtsev
\textit{et al}. \citep{2006}. The key is that our protocol is heralded.
The successful generation of $\vert\Phi^{\prime}\rangle$ is signaled
by a click at the idler detector ($D_{1}$). This means the experiment
can be run continuously and at high repetition rates (e.g., with a
pulsed laser). For instance, even with a success probability as low
as $10^{-5}$, a $10$$MHz$ to $1$$GHz$ repetition rate \citep{Ma_2015,XUE2024107048,CHEN2025111703}
yields about $10^{2}\thicksim10^{4}$ heralded states per second.
For many proof-of-principle quantum information protocols including
QKD, this rate is entirely practical.

Having demonstrated nG state generation from coherent inputs, we now
apply our protocol to a SV input. We will show that this process generates
a state equivalent to a squeezed photon number state. In the previous
subsection, we considered a coherent state as the signal input and
demonstrated that the resulting output states exhibited nG characteristics.
In this subsection, we consider another typical example, a single-mode
SV state. The SV state is defined as $\vert r\rangle=S(r)\vert0\rangle$,
where the squeezing operator is given by $S(r)=\exp\left[\frac{r}{2}(b^{2}-b^{\dagger2})\right]$,
and $r$ is the squeezing parameter. For $r>0$, the SV state is squeezed
along the position quadrature $x=\frac{1}{\sqrt{2}}(b+b^{\dagger})$
and anti-squeezed in the momentum quadrature $p=\frac{-i}{\sqrt{2}}(b-b^{\dagger})$;
the reverse holds for $r<0$. The squeezed vacuum state is a fundamental
resource in quantum optics \citep{PhysRevA.71.055801}, with substantial
interest owing to its distinctive properties and wide-ranging applications
in quantum sensing \citep{PhysRevLett.68.3020,2011}, quantum communication
\citep{RevModPhys.77.513}, and others {[}see, e.g., Refs.~\citep{PhysRevLett.117.110801,Andersen2016,SCHNABEL20171}{]}.
In addition, the photon-added and photon-subtracted SV states have
been studied extensively \citep{2007PV}. In general, generating multiphoton-added
or multiphoton-subtracted SV states suffers from very low success
probabilities due to measurement imperfections and operational inefficiencies.
From an implementation perspective, photon subtraction is typically
easier than photon addition. Therefore, the preparation of multiphoton-added
SV states requires advanced techniques and higher efficiency. Herein,
we propose a simple method to generate two-photon-added squeezed vacuum
(TPASV) states without requiring actual photon-addition operations.
In the schematic of our setup, if the initial signal (pointer) state
is prepared as an SV state, the final signal output state after the
measurement process can be expressed as 
\begin{align}
|\varPsi\rangle & =\chi S(r)\left[|0\rangle+\mu|2\rangle\right]\label{eq:29}
\end{align}
where the normalization coefficient $\chi$ equals to $\chi=(1+\mu^{2})^{-1/2}$
with

\begin{equation}
\mu=\frac{i\frac{g}{2\sqrt{2}}\left(1+\frac{1}{\epsilon}\right)sinh\left(2r\right)}{1-i\frac{g}{2}\left(1+\frac{1}{\epsilon}\right)sinh^{2}\left(r\right)}.\label{eq:32}
\end{equation}
The signal output state is a superposition of a SV state and a squeezed
two-photon state. We can express $\vert\Psi\rangle$ in a more compact
form as 
\begin{align}
\vert\Psi\rangle & =\eta\left[|r\rangle-ig\sinh^{2}r\langle n_{a}\rangle_{w}\mathcal{N}_{b^{\dagger2}}^{-1}\vert\psi_{b^{\dagger2}}\rangle\right],\label{eq:34}
\end{align}
where the normalized expression of the TPASV state $b^{\dagger2}\vert r\rangle$
is defined by 
\begin{equation}
\vert\psi_{b^{\dagger2}}\rangle=\mathcal{N}_{b^{\dagger2}}S\left(r\right)\left[\vert0\rangle-\sqrt{2}\left(\tanh r\right)^{-1}\vert2\rangle\right]\label{eq:18}
\end{equation}
with normalization constant $\mathcal{N}_{b^{\dagger2}}=[1+2(\tanh r)^{-2}]^{-1/2}$.
The normalization constant $\eta$ for the state $\vert\Psi\rangle$
is given by 
\begin{equation}
\eta^{-2}=1+g^{2}\langle n_{a}\rangle_{w}^{2}\ sinh^{4}(r)+\frac{g^{2}}{2}\langle n_{a}\rangle_{w}^{2}\ sinh^{2}(2r).
\end{equation}
 For the parameters considered herein, the overall success probability
of generating the state $\vert\Psi\rangle$ is approximately $10^{-4}$,
as determined by the expression $\epsilon^{2}\vert\mathcal{\eta\vert}^{-2}$. 

\begin{figure}
\includegraphics[width=8cm]{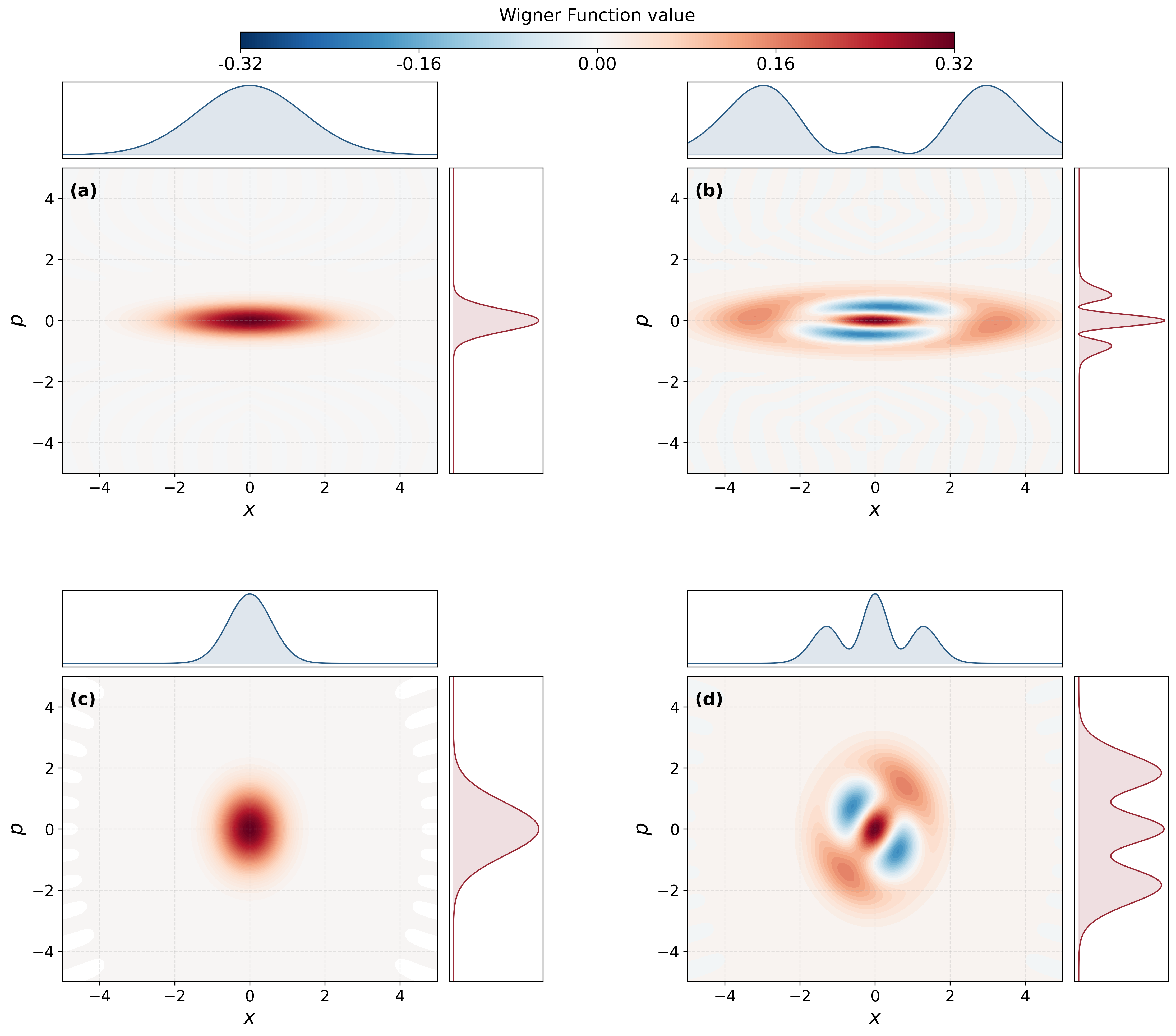}

\caption{\label{fig:4} \textbf{Wigner functions (contour plots) and quadrature
distributions for the generated state $\vert\Psi\rangle$.} \textbf{a}
and \textbf{c} show the Wigner functions of the signal input SV states
with squeezing parameters $r=-0.7$ and $r=0.2$, respectively. The
corresponding Wigner functions of the output signal states $\vert\Psi\rangle$
are shown in \textbf{b} and \textbf{d}. Other parameters are the same
as those used in Fig.~\ref{fig:2}.}
\end{figure}

The TPASV state is generated by successively adding two photons to
the SV state. Interestingly, this process is also equivalent to subtracting
one photon from the SV state, and subsequently adding one photon.
That is, 
\begin{equation}
b^{\dagger2}\vert r\rangle=-\coth r\ b^{\dagger}b\vert r\rangle.\label{eq:36}
\end{equation}
It follows that the normalized TPASV state is identical to the normalized
photon-subtracted-and-added squeezed state: $\vert\psi_{b^{\dagger2}}\rangle=\vert\psi_{b^{\dagger}b}\rangle$.
This relation can be verified using the following operator identities:
\begin{subequations}
\begin{align}
b^{\dagger}S\left(r\right)\vert0\rangle & =\cosh rS\left(r\right)\vert1\rangle,\label{eq:37}\\
bS\left(r\right)\vert0\rangle & =-\sinh rS\left(r\right)\vert1\rangle,\label{eq:38}
\end{align}
 \end{subequations}along with the unitary transformations: \begin{subequations}
\begin{align}
S^{\dagger}\left(r\right)bS\left(r\right) & =b\cosh r-b^{\dagger}\sinh r,\\
S^{\dagger}\left(r\right)b^{\dagger}S\left(r\right) & =b^{\dagger}\cosh r-b\sinh r.
\end{align}
 \end{subequations}From Eqs.~(\ref{eq:37}) and (\ref{eq:38}),
we see that the squeezed single-photon state $S(r)\vert1\rangle$
can be obtained by adding or subtracting a single photon to or from
an SV state. As confirmed in previous studies \citep{PhysRevA.70.020101,2006},
the SV state and squeezed single-photon state are good approximations
of small-amplitude even and odd SC states, respectively. The SC states
are defined as 
\begin{equation}
\vert SCS_{\pm}(\alpha)\rangle=\mathcal{N}_{\pm}\left(\vert\alpha\rangle\pm\vert-\alpha\rangle\right),\label{eq:25}
\end{equation}
where the normalization constants are given $\mathcal{N}_{\pm}=(2\pm2e^{-2\alpha^{2}})^{-1/2}$.
When expressed on the Fock basis, the even SC state contains only
even photon-number components, whereas the odd SC state consists only
of odd-number components.

In our scheme, the second term in the signal output state is $\vert\psi_{b^{\dagger2}}\rangle$
{[}see Eq. (\ref{eq:18}){]}, which is proportional to the weak value
$\langle n_{a}\rangle_{w}$. As discussed previously, when $\langle n_{a}\rangle_{w}$
has a large value, the second term becomes dominant in the output
signal state {[}see Eq. (\ref{eq:34}){]}, enabling us to effectively
obtain the TPASV state without performing actual photon-addition operations.

As noted above, the generated state $\vert\Psi\rangle$ takes the
form $c_{0}\vert0\rangle+c_{2}\vert2\rangle$, with $c_{0},c_{2}\in\mathbb{C}$.
That is, it is a squeezed superposition of the vacuum and two-photon
Fock states. Such a state provides an excellent approximation of a
squeezed even SC state with large amplitude $\alpha$ \citep{PhysRevA.78.063811},
which is defined as 
\begin{equation}
\vert\Psi_{sscs}\rangle=\mathcal{N}_{sscs}S\left(r^{\prime}\right)\left[\vert\alpha\rangle+\vert-\alpha\rangle\right],\label{eq:42}
\end{equation}
where the normalization constant is $\mathcal{N}_{sscs}=\left(2+2e^{-2\alpha^{2}}\right)^{-1/2}$.
This approximation is valid because the squeezing operator $S(r')$,
with an appropriately chosen squeezing parameter $r'$, effectively
pulls apart the two peaks of the cat state. To explain this effect
more clearly, in Fig. \ref{fig:4}, we plotted the Wigner functions
of our input SV and the corresponding signal output state $\vert\Psi\rangle$
for different squeezing parameters $r$. Using our protocol, the input
Gaussian-type SV state transformed to the nG state characterized by
Wigner negativity, and squeezing occurred along both $x$ and $p$
quadratures {[} see Figs. \ref{fig:4} \textbf{b} and \textbf{d} {]}.
The interference fringes and quadrature distribution features of the
state $\vert\Psi\rangle$ showed in Fig. \ref{fig:4} is very similar
to the squeezed even SC state which is the basis of GKP states \citep{PhysRevA.64.012310,PhysRevLett.128.170503}.
It also indicates that the two peaks of the generated state are separated
along the $x$ or $p$ direction depending on the value of the squeezing
parameter $r$. In the following, we further analyze the similarity
between our generated state $\vert\Psi\rangle$ and the ideal large-amplitude
squeezed even  SC state using fidelity and phase-space distributions.
In the following, we assume that both squeezing parameters $r$ and
$r'$ are real.

The fidelity $F_{2}$ between the generated state $\vert\Psi\rangle$
and the ideal squeezed even SC state $\vert\Psi_{sscs}\rangle$ is
given by 
\begin{align}
F_{2} & =\vert\langle\Psi_{sscs}\vert\Psi\rangle\vert^{2}\nonumber \\
 & =\frac{4\mu e^{\frac{-2\alpha^{2}}{1+\mu^{2}}}}{1+\mu^{2}}\left|\chi\mathcal{N}_{sscs}\left(\sqrt{2}+\mu^{*}\frac{1-\nu^{4}+4\nu^{2}\alpha^{2}}{1+\nu^{2}}\right)\right|^{2},\label{eq:43}
\end{align}
where $\nu=e^{-(r-r^{\prime})}$.

\begin{figure}
\includegraphics[width=8cm]{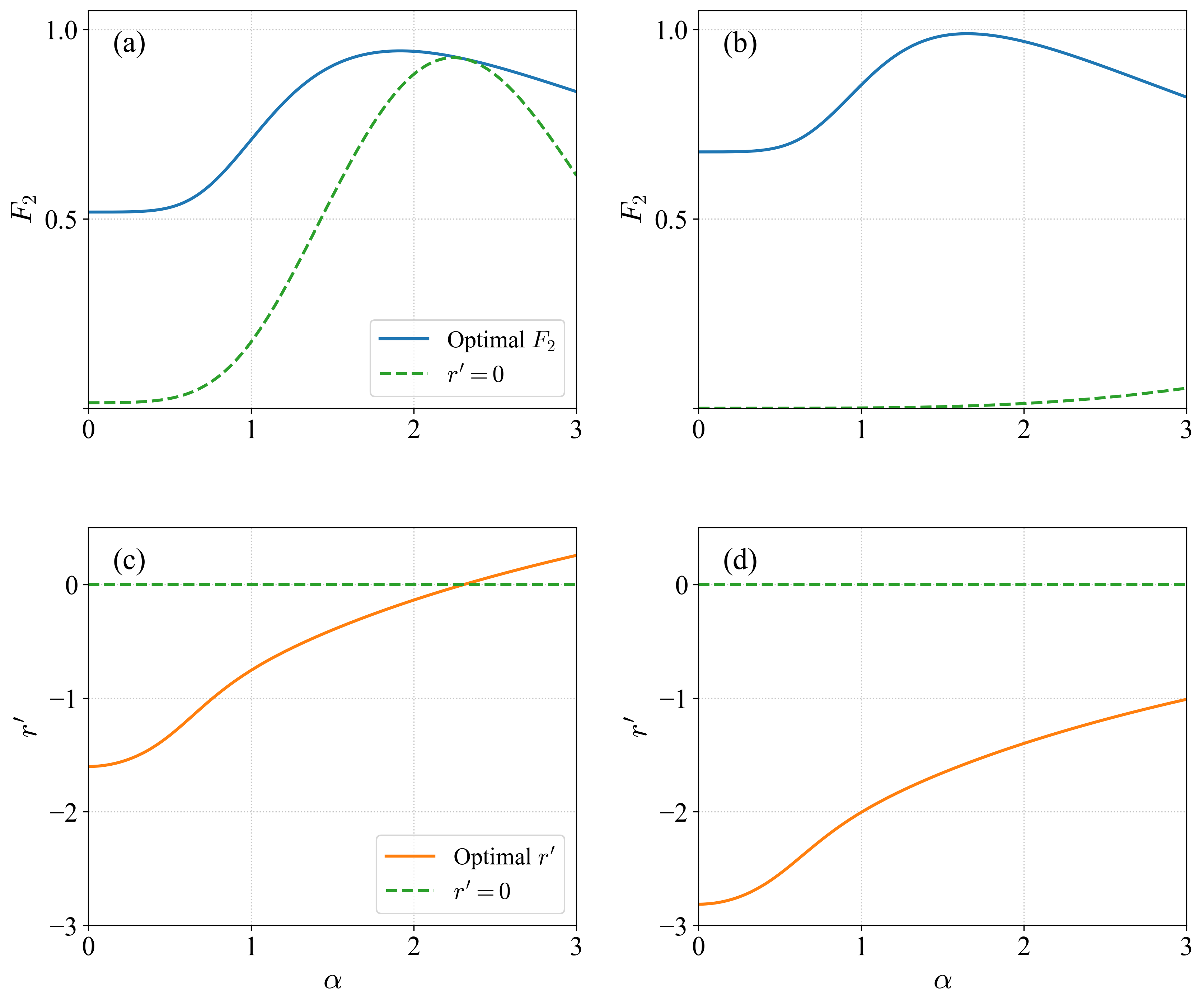} \caption{\label{fig:10} \textbf{Optimal fidelities and squeezing parameters.}
\textbf{a} and \textbf{b }show the optimal fidelities between the
output signal state $\vert\Psi\rangle$ and the target squeezed even
SC state (solid curves), as well as the ideal even SC state (dashed
curves), for different system parameters. \textbf{c} and \textbf{d}
show the corresponding squeezing parameter $r^{\prime}$ of the target
squeezed even SC state for which the fidelities in \textbf{a} and
\textbf{b} are optimized. The squeezing parameters of the initial
SV states are $r=-0.7$ and $r=-2$ for \textbf{a} and \textbf{b},
respectively. The other parameters are the same as those used in Fig.~\ref{fig:2}.}
\end{figure}

In Fig.~\ref{fig:10}, we present the optimal fidelity between $\vert\Psi\rangle$
and $\vert\Psi_{sscs}\rangle$ as a function of the coherent amplitude
$\alpha$ for different squeezing parameter $r$. In these plots,
for each value of $\alpha$, we optimize the squeezing parameter $r'$
of the ideal squeezed even SC state to maximize its overlap with the
generated state $\vert\Psi\rangle$. The cases $r=-0.7$ and $r=-2$
are shown in panels \textbf{a} and \textbf{b}, respectively. The corresponding
optimized values of $r'$ as a function of $\alpha$ are plotted in
panels (c) and (d), respectively.

In Figs. \ref{fig:10}\textbf{a} and \ref{fig:10}\textbf{b}, the
blue solid curves represent the optimal fidelity between $\vert\Psi\rangle$
and $\vert\Psi_{sscs}\rangle$, while the green dashed curves show
the fidelity between $\vert\Psi\rangle$ and the ideal even SC state
(without squeezing). As shown in Fig. \ref{fig:10}\textbf{a} for
$r=-0.7$, the optimal fidelity reaches as high as $F_{2}=0.97$ at
$\alpha=1.91$ and $r'=-0.14$, indicating a high degree of similarity
between the generated state and the ideal squeezed even SC state.
As shown in Fig.~\ref{fig:10}\textbf{b}, when $r=-2$, the output
signal state $\vert\Psi\rangle$ provides an excellent approximation
to the squeezed even SC state, achieving a very high fidelity of $F_{2}=0.994$
at $\alpha=2.07$ and $r'=-0.192$. This result indicates that, in
this case, the output signal state $\vert\Psi\rangle$ is extremely
well approximated by the ideal squeezed even SC state. 

\begin{figure}[H]
\includegraphics[width=8cm]{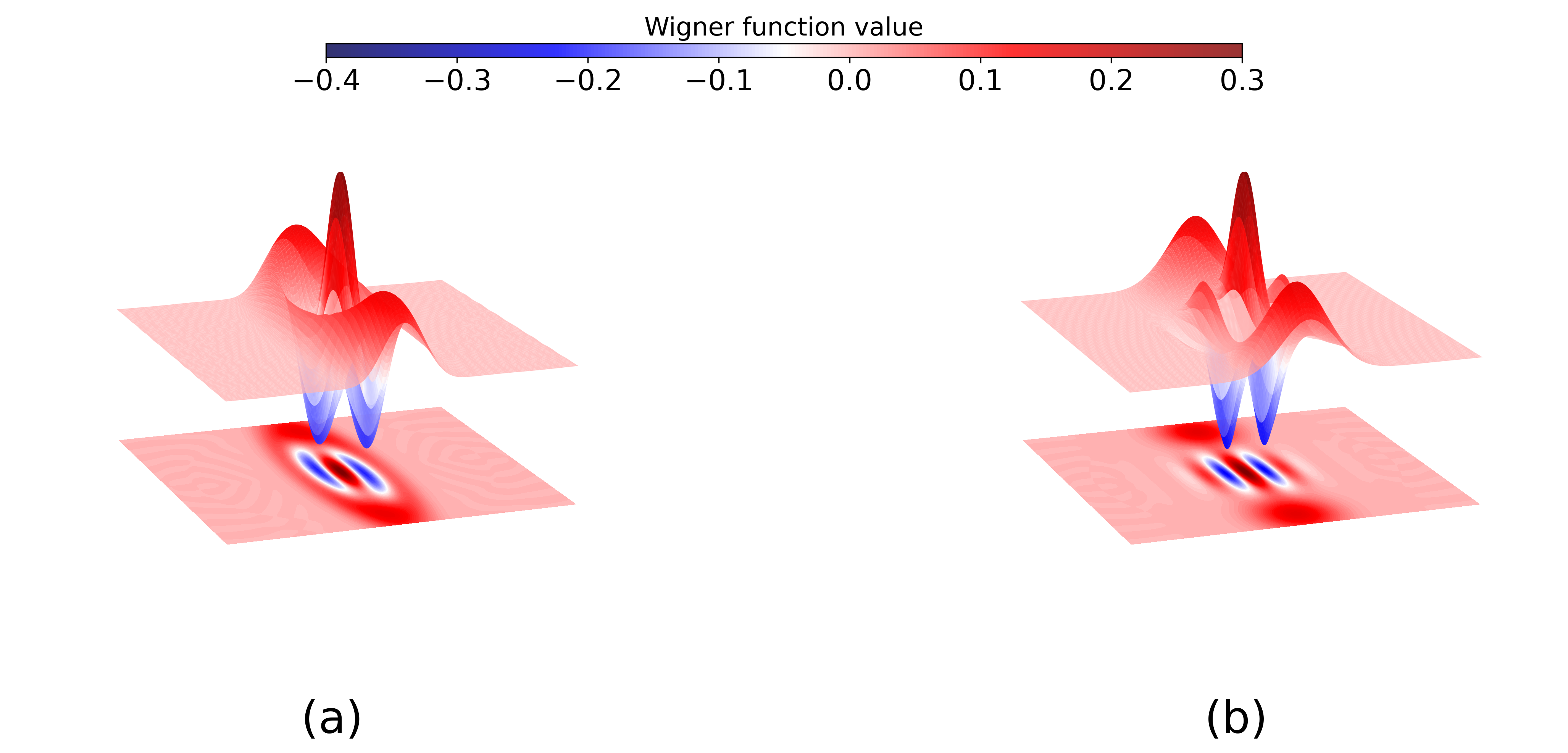} \caption{\label{fig:12} \textbf{Wigner functions of the generated state $\vert\Psi\rangle$.}
\textbf{a} Wigner function of the generated state $\vert\Psi\rangle$
with $r=-0.7$. \textbf{b} Wigner function of the ideal squeezed even
SC state with $r^{\prime}=-0.14$ and $\alpha=2$. Other parameters
are the same as those used in Fig.~\ref{fig:2}. For these parameters,
the fidelity calculated using the Wigner functions of the generated
and ideal squeezed even SC states is $F_{2}\approx0.97$.}
\end{figure}

While fidelity provides a useful quantitative measure of the similarity
between the generated state and the target state, it is also informative
to examine the Wigner functions, as each quantum state exhibits unique
features in its phase-space distribution. To this end, in Fig.~\ref{fig:12},
we present the Wigner functions of the output signal state $\vert\Psi\rangle$
and compare them with those of the ideal squeezed even SC state $\vert\Psi_{sscs}\rangle$. 

Figure~\ref{fig:12}\textbf{a} shows the Wigner function of $\vert\Psi\rangle$
for $r=-0.7$, while Fig. \ref{fig:12}\textbf{b} displays the Wigner
function of the squeezed even SC state with $r'=-0.14$ and $\alpha=2$.
For these parameters, the fidelity between the two states is approximately
$F_{2}\approx0.97$. This further confirms the high similarity of
our generated output signal state $\vert\Psi\rangle$ to the $\vert\Psi_{sscs}\rangle$
with large amplitude $\alpha$. Fig. \ref{fig:12} also illustrates
the effects of our protocol on the initially prepared Gaussian SV
state, which transforms it into a nG state.

As mentioned previously, the generation of SC states with high fidelity
($F\geq0.99$) while maintaining large coherent amplitudes ($\alpha\geq2$)
is a crucial task. Although our scheme does not directly generate
large-amplitude SC states, the output signal state achieves exceptionally
high fidelity with a squeezed even SC state possessing a large amplitude
$\alpha$. Therefore, our approach could be applied to problems where
large-amplitude SC states are required, offering potentially better
performance. \vspace{0.3cm}

\textbf{Enlargement of Schr\"{o}dinger Cat States}

In the preceding discussion, we demonstrated the usefulness of our
scheme in generating nG states from Gaussian input states. We now
explore the advantages of our protocol for amplifying SC states. In
particular, we investigate the possibility of enhancing the non-Gaussianity
of nG states by taking SC states as an example.

If the input signal state is one of the SC states defined in Eq.~(\ref{eq:25}),
the output signal state produced by our scheme takes the following
form: 
\begin{align}
\vert\Psi_{\pm}\rangle & =\mathcal{N_{\pm}^{\prime}}\left[\left(1-i\kappa^{\prime}b^{\dagger}b\right)\vert\alpha\rangle\pm\left(1+i\kappa^{\prime}b^{\dagger}b\right)\vert-\alpha\rangle\right]\nonumber \\
 & =\mathcal{N_{\pm}^{\prime}}\left[\vert\alpha\rangle\pm\vert-\alpha\rangle-i\kappa^{\prime}b^{\dagger}b(\vert\alpha\rangle\pm\vert-\alpha\rangle)\right]\label{eq:28-1}
\end{align}
where $\kappa^{\prime}=g\langle n_{a}\rangle_{w}$ and $\mathcal{N}_{\pm}^{\prime}$
is the normalization constant, given by 
\begin{equation}
\mathcal{N_{\pm}^{\prime}}=\frac{1}{\sqrt{2}}\left[\left(\kappa^{\prime2}\alpha{}^{4}+1\right)\left(1\pm e^{-2\alpha^{2}}\right)+\kappa^{\prime2}\alpha{}^{2}\left(1\mp e^{-2\alpha^{2}}\right)\right]^{-\frac{1}{2}}.\label{eq:28-2}
\end{equation}
The success probability for this process is given by $\epsilon^{2}\vert\mathcal{N_{\pm}^{\prime}}\vert^{-2}$.
For the parameters considered here ($\epsilon\thicksim10^{-3}$, $g=0.01$),
this results in probabilities in the range of $10^{-6}$ to $10^{-3}$,
with higher fidelities typically corresponding to the more favorable
end of this range.

As seen from Eq.~(\ref{eq:28-1}), the output state is a superposition
of the initial SC state and photon-number-weighted SC state $b^{\dagger}b\vert SCS_{\pm}\rangle$.
Since the weak value $\langle n\rangle_{w}$ can become large in our
protocol, the second term can dominate, effectively transforming the
initial SC state into $b^{\dagger}b\vert SCS_{\pm}\rangle$ via the
WVA.

Although the output state $\vert\Psi_{\pm}\rangle$ differs slightly
in form from an ideal SC state, the deviation is relatively small.
To quantitatively evaluate the similarity, we computed the fidelity
between $\vert\Psi_{\pm}\rangle$ and an ideal SC state with a potentially
different coherent amplitude $\alpha^{\prime}$, defined as 
\begin{align}
F_{\pm} & =\vert\langle SCS_{\pm}(\alpha^{\prime})\vert\Psi_{\pm}\rangle\vert^{2}\nonumber \\
 & =4\left|\mathcal{N}_{\pm}\mathcal{N}_{\pm}^{\prime}\left[\left(1-i\kappa^{\prime}\alpha^{\prime}\alpha\right)h_{-}\pm\left(1+i\kappa^{\prime}\alpha^{\prime}\alpha\right)h_{-}\right]\right|^{2},\label{eq:29-1}
\end{align}
where $h_{\pm}=\exp\left[-\frac{1}{2}|\alpha'\pm\alpha|^{2}\right]$.
In the above calculations, we assume that both $\alpha$ and $\alpha'$
are real.

\begin{figure}
\includegraphics[width=8cm]{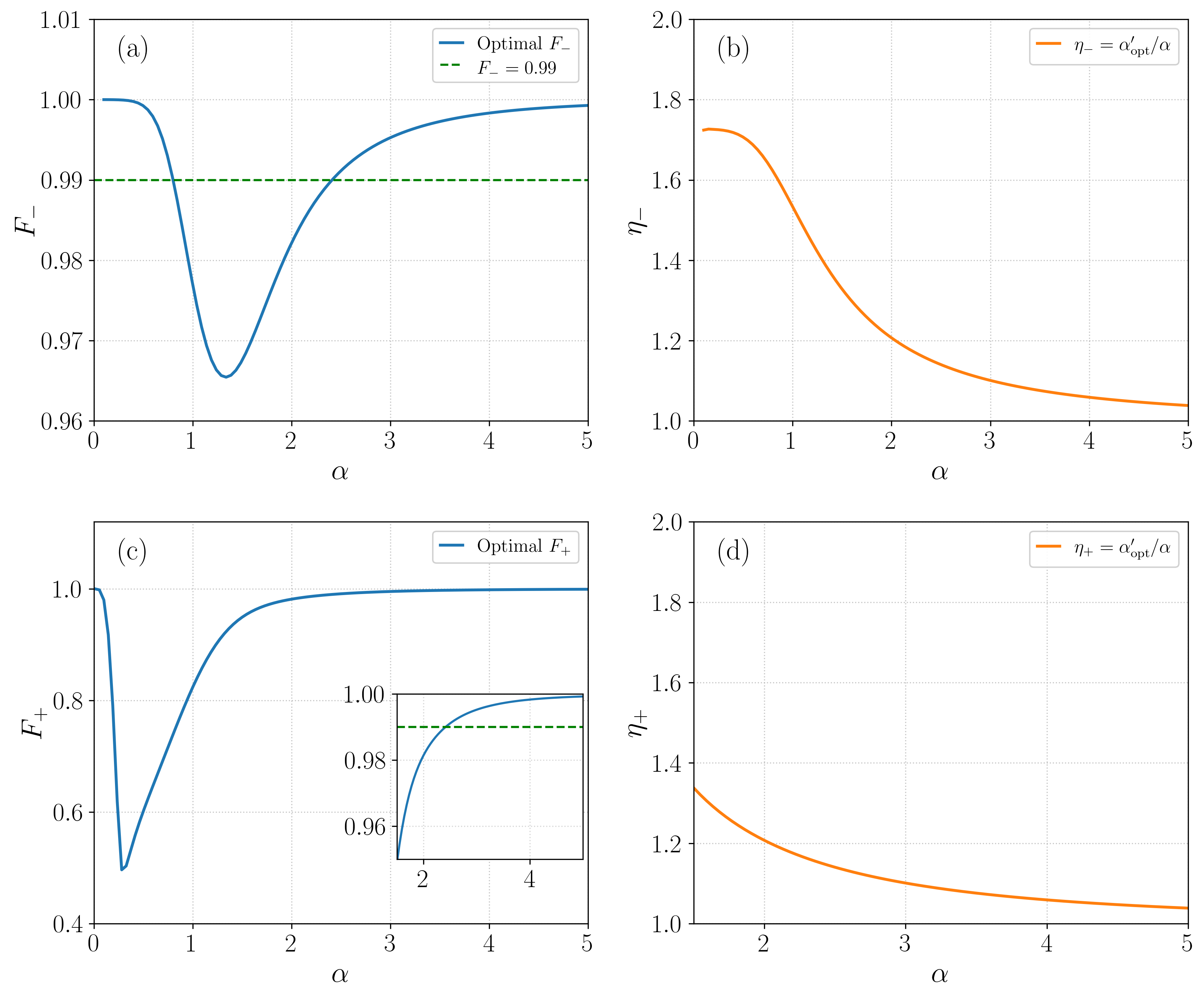}

\caption{\label{fig:11} \textbf{Optimal fidelities and enlargement coefficients.}
\textbf{a} and \textbf{c} show the optimal fidelities between the
output signal states $\vert\Psi_{\pm}\rangle$ and the ideal SC states
$\vert SCS_{\pm}(\alpha^{\prime})\rangle$, plotted as a function
of the coherent state amplitude $\alpha$. \textbf{b} and \textbf{d}
show the corresponding enlargement coefficients $\eta_{-}=\alpha^{\prime}/\alpha$
and $\eta_{+}=\alpha^{\prime}/\alpha$, for the odd and even SC states,
respectively, for which the fidelities in \textbf{a} and \textbf{c}
are optimized. The other parameters are the same as those used in
Fig.~\ref{fig:2}.}
\end{figure}

In Fig.~\ref{fig:11}, we show the optimal fidelity functions $F_{+}$
and $F_{-}$ between our generated output states $\vert\Psi_{+}\rangle$
and $\vert\Psi_{-}\rangle$ and the ideal even and odd SC states,
respectively, as functions of coherent state amplitude $\alpha$.
The corresponding optimized amplitudes $\alpha'$ of the ideal SC
states are used for the maximal overlap. As shown in Fig.~\ref{fig:11}\textbf{a},
the generated state $\vert\Psi_{-}\rangle$ provides an excellent
approximation to the ideal odd SC state, with fidelity $F_{-}\geq0.965$
across a wide range of amplitudes, $\alpha$. Notably, in the regions
$0<\alpha<0.8$ and $\alpha\geq2.45$, the fidelity exceeds 0.99.
For the even SC state input case {[}Fig.~\ref{fig:11}\textbf{c}{]},
the generated state $\vert\Psi_{+}\rangle$ initially shows noticeable
deviation from the ideal even SC state in the small-amplitude regime
($0<\alpha<0.29$). However, as $\alpha$ increases beyond 0.29, the
fidelity $F_{+}$ improves significantly, surpassing $0.9$ for all
$\alpha>1.24$ regions. Thus, for $\alpha>1.24$, $\vert\Psi_{+}\rangle$
also can be considered an excellent approximation to the ideal even
SC state. Moreover, in high-fidelity regions, we observed that the
optimized coherent state amplitudes $\alpha'$ of the ideal SC states
are consistently larger than the corresponding values of $\alpha$
for our generated states $\vert\Psi_{\pm}\rangle$ {[}see Figs.~\ref{fig:11}\textbf{b}
and \ref{fig:11}\textbf{d}{]}. This indicates that our protocol effectively
enlarges the SC states. This enlargement effect can be further confirmed
by comparing the Wigner functions of the input ideal SC states and
the generated output states $\vert\Psi_{\pm}\rangle$.

\begin{figure}
\includegraphics[width=8cm]{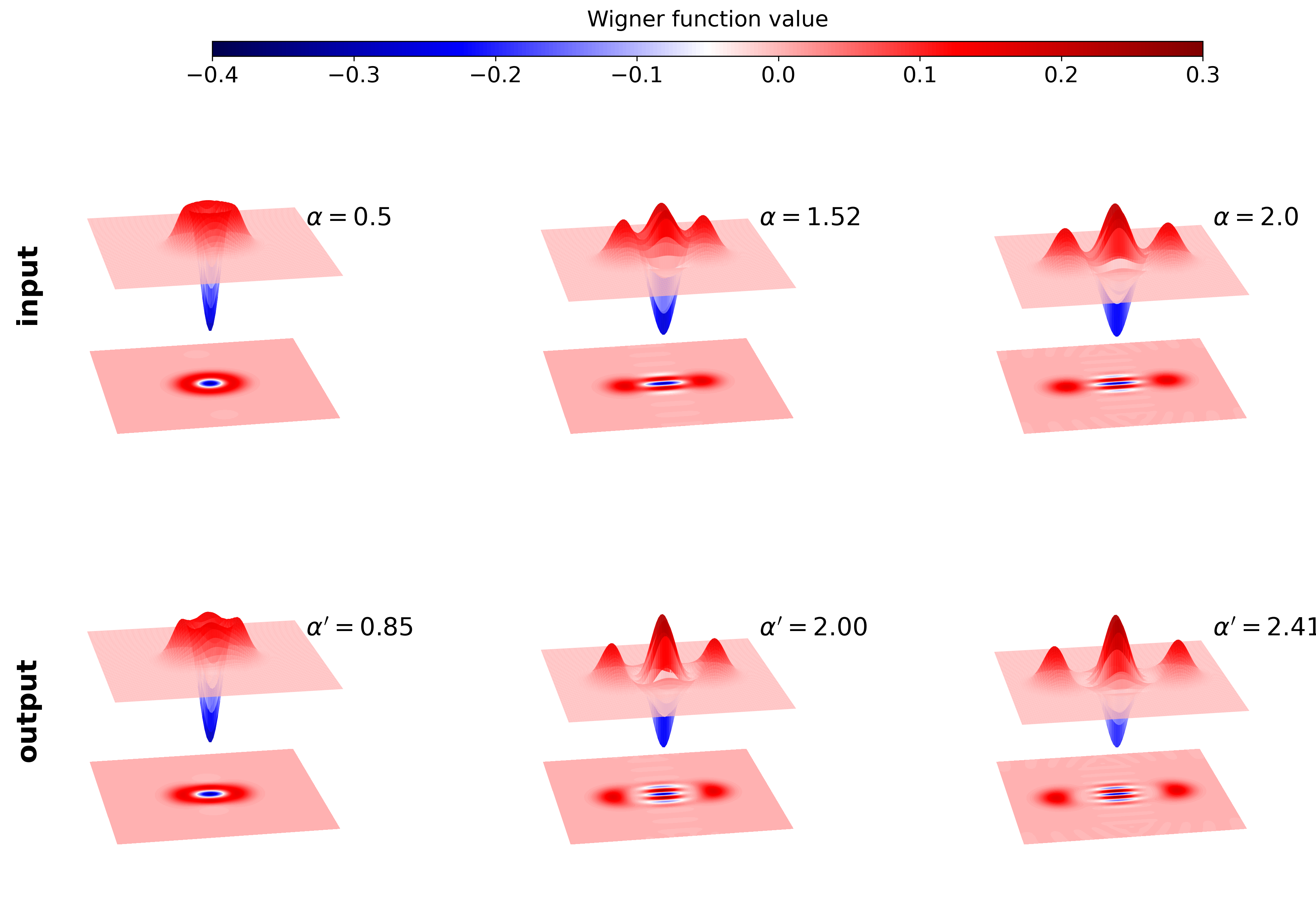}

\caption{\label{fig:8-1} \textbf{Wigner functions of odd SC states.} The first
row presents the Wigner functions of the input odd SC states for different
coherent amplitudes $\alpha$. The second row shows the corresponding
Wigner functions of the generated states $\vert\Psi_{-}\rangle$ for
optimized amplitudes $\alpha^{\prime}$, chosen to maximize the fidelity
with the ideal odd SC state $\vert SCS_{-}(\alpha)\rangle$. Other
parameters are the same as those used in Fig.~\ref{fig:2}.}
\end{figure}

In the small-amplitude regime ($\alpha\leq1.2$), the odd SC state
is well approximated by the squeezed single-photon state $S(r)\vert1\rangle$,
with a fidelity exceeding $0.99$ \citep{2004,2006}. The corresponding
optimized squeezing parameter that maximizes the fidelity for a given
$\alpha$ is given by $r_{opt}(\alpha)=\ln\left(\sqrt{\frac{2\alpha^{2}}{3}+\frac{1}{3}\sqrt{9+4\alpha^{2}}}\right)$.
However, for larger amplitudes ($\alpha\geq2$), the fidelity rapidly
decreased below $87.8\%$. In contrast, we find that our protocol
can generate odd SC states with high fidelity, even for large amplitudes
$\alpha\geq2$. In Fig.~\ref{fig:8-1}, we plot the Wigner functions
of the odd SC input states for different values of $\alpha$, along
with the corresponding output states $\vert\Psi_{-}\rangle$, using
the optimized amplitude $\alpha^{\prime}$ which maximizes the fidelity
between $\text{\ensuremath{\vert}\ensuremath{\Psi_{-}\rangle}}$ and
the ideal odd SC target state $\vert SCS_{-}(\alpha)\rangle$. As
shown, for each input odd SC state, our protocol yields an excellent
approximation to a larger-amplitude odd SC state encoded in the output
state $\vert\Psi_{-}\rangle$. For example, when the input amplitudes
are $\alpha=0.5$, $1.52$, and $2.0$, the corresponding enlarged
SC states, characterized by the output amplitudes $\alpha'=0.85$,
$2.00$, and $2.41$, are obtained with high fidelities: $F_{-}=0.999$,
$0.968$, and $0.982$, respectively. The amplitude enlargement ratio
$\eta_{-}=\alpha^{\prime}/\alpha$ is plotted in Fig.~\ref{fig:11}\textbf{b}.
The curve shows that $\eta_{-}>1$ across a wide range of $\alpha$,
indicating that our protocol effectively amplifies the amplitude of
a given odd SC state while maintaining high fidelity. 
\begin{figure}
\includegraphics[width=8cm]{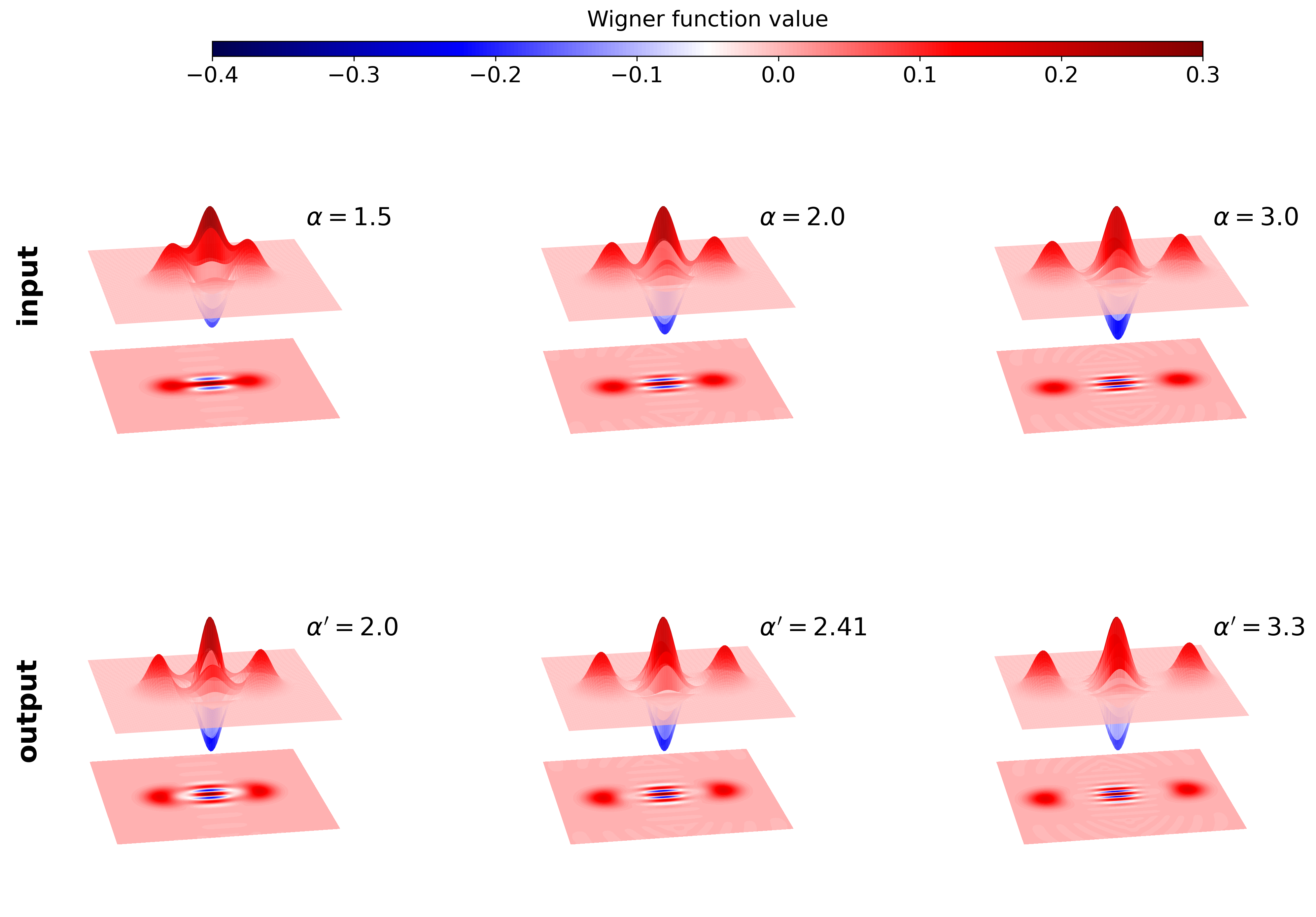}

\caption{\label{fig:9} \textbf{Wigner functions of even SC states.} The first
row presents the Wigner functions of the input even SC states for
different coherent amplitudes $\alpha$. The second row shows the
corresponding Wigner functions of the generated states $\vert\Psi_{+}\rangle$
for optimized amplitudes $\alpha^{\prime}$, chosen to maximize the
fidelity with the ideal even SC state $\vert SCS_{+}(\alpha)\rangle$.
Other parameters are the same as those used in Fig.~\ref{fig:2}.}
\end{figure}

The amplitude-enlarging advantage of our protocol for SC states can
also be confirmed by using the even SC state as an example. Similar
to the odd SC case, our protocol can generate large-amplitude even
SC states with high fidelity when a small-amplitude even SC state
is used as input. As shown in Fig.~\ref{fig:9}, when the input even
SC states have amplitudes $\alpha=1.5$, $2.0$, and $3.0$, the corresponding
enlarged even SC states with amplitudes $\alpha^{\prime}=2.0$, $2.41$,
and $3.3$ are obtained, with fidelities $F_{+}=0.950$, $0.981$,
and $0.995$, respectively. These enlarged states are encoded in the
output state $\vert\Psi_{+}\rangle$, generated by our protocol for
a given ideal small-amplitude input even SC state. As shown in Fig.~\ref{fig:11}\textbf{d},
the amplitude enlargement ratio $\eta_{+}=\alpha^{\prime}/\alpha$
for even SC states remains greater than one for all tested values
of $\alpha$, further confirming that our protocol exhibits a robust
amplitude enhancing effect for even SC states. Moreover, the even
SC state closely resembles a Gaussian squeezed vacuum state when $\alpha\lessapprox0.75$,
and in that regime, it serves as a high-fidelity ($F=0.99$) approximation
to the SV state \citep{PhysRevA.78.052304}. Therefore, for even SC
inputs with $\alpha\lessapprox0.75$, the output behavior mirrors
the earlier analysis for squeezed vacuum inputs, using the optimized
squeezing parameter $r_{opt}(\alpha)=\ln\left(\sqrt{2\alpha^{2}+\sqrt{1+4\alpha^{2}}}\right)$.
Although the even SC state is approximately Gaussian in this small-amplitude
regime, the corresponding output state generated by our protocol displays
strong nG features, as indicated by the appearance of Wigner negativity
{[}see Figs.~\ref{fig:4}\textbf{c} and \ref{fig:4}\textbf{d}{]}.
The amplitude enlargement effect of our protocol can also be directly
confirmed by comparing the average photon numbers of the input and
output states. Large-amplitude SC states naturally contain more photons
than small-amplitude states. In the output states $\vert\Psi_{\pm}\rangle$,
the first term corresponds to the input SC state, whereas the second
term, $b^{\dagger}b\vert SCS_{\pm}\rangle$, contributes significantly
to the photon number. As a result, the average photon number in the
output states $\vert\Psi_{\pm}\rangle$ is always larger than that
in the corresponding ideal SC inputs, while still maintaining high
fidelity within the discussed parameter regimes.

It should be emphasized that the amplitude-enlarging feature of our
protocol for SC states remains valid across a wide range of coherent
amplitudes $\alpha$, with the corresponding fidelity $F_{-}$ ($F_{+}$)
exceeding 0.99 for $\alpha\gtrsim2.41$. In these regimes, the enlargement
coefficients $\eta_{-}$ ($\eta_{+}$) are also consistently greater
than one. The above results indicate that even though we did not directly
prepare the exact large-amplitude SC states, the equivalent SC states
$\vert SC_{\pm}(\alpha^{\prime})\rangle$ were encoded in our signal
output states $\vert\Psi_{\pm}\rangle$. Consequently, the two coherent
components $\vert\pm\alpha^{\prime}\rangle$ of the equivalent SC
states $\vert SC_{\pm}(\alpha^{\prime})\rangle$ become approximately
orthogonal when $\alpha^{\prime}\ge2$, making such states valuable
resources for exploring macroscopic quantum superpositions and quantum
information applications. However, to maintain the weak nonlinearity
condition $g\langle n_{b}\rangle\ll1$, the coherent amplitudes $\alpha$
of the initial input SC states cannot be too large.

As mentioned in the Introduction, optical SC states can be generated
via cross-Kerr interactions characterized by third-order nonlinear
susceptibility $\chi^{(3)}$ \citep{PhysRevA.59.4095,Int}. However,
in practice, generating large-amplitude SC states under realistic
conditions is extremely challenging due to the typically small and
attenuated values of $\chi^{(3)}$. As analyzed above, by using the
WVA technique to effectively enhance single-photon level nonlinearities,
our protocol enables the generation of large-amplitude SC states ($\alpha\geq2$)
with high fidelity, even in the presence of weak cross-Kerr interactions.

In recent years, several methods have been explored for enlarging
SC states \citep{PhysRevA.70.020101,2017n,2019R,2024Li}. Among these,
the synthesis-based method for enlarging even SC states, as described
in Refs. \citep{PhysRevA.70.020101,2017n} requires the input to be
in a small odd SC state. However, this approach is constrained by
the amplitude of the initial state and is resource-intensive, making
the generation of large-amplitude SC states experimentally demanding.
In Ref. \citep{2019R}, SC states with amplitudes up to $\alpha=1.4$
were generated through direct interaction between a coherent light
pulse and a single-sided cavity containing a three-level atom. Although
promising, this method is limited in scalability to larger amplitudes,
and is susceptible to decoherence when strong light fields are used.

In contrast, our protocol provides a simple and experimentally feasible
alternative for generating large-amplitude SC states with high fidelity
across a broad range of amplitudes $\alpha$, without requiring explicit
photon addition or subtraction operations. However, we note that the
success probability of our scheme is not high and is proportional
to the square of the beam splitter deviation from the ideal $50:50$
configuration, i.e., $\epsilon{}^{2}$. \vspace{0.3cm}

\textbf{DISCUSSION }

In this study, we have proposed a protocol for generating nG states
by combining postselected WM with a weak cross-Kerr interaction. The
core mechanism relies on injecting a single photon into an interferometer
through the idler input port and heralding a successful event by its
detection at a dark port. This process amplifies the single-photon
nonlinearity via WVA, effectively transforming a given input signal
state into a nG output without explicit photon-addition or -subtraction
hardware.

Our theoretical analysis demonstrates that this versatile framework
can generate a wide range of high-fidelity nG states. With coherent
state inputs, the protocol produces states closely resembling SPAC
states. For SV inputs, it yields high-fidelity TPASV states and their
equivalents, squeezed number states. A particularly significant result
is the protocol's ability to act on nG inputs themselves: when applied
to SC states, it not only enhances their non-Gaussianity, as evidenced
by increased Wigner negativity, but also effectively amplifies small-amplitude
cat states into large-amplitude ones ($\alpha\geq2$) while maintaining
fidelities exceeding 0.99, particularly in the case of odd SC states.

The primary advantage of our protocol is its versatility; a single
apparatus can generate a continuous family of nG states, contrasting
with methods that require specialized setups for each state type \citep{2004,2006}.
Its ability to amplify SC states addresses a key challenge in quantum
optics, offering a distinct pathway compared to conventional photon
operations that often degrade state quality at high amplitudes \citep{PhysRevLett.101.233605,PhysRevA.110.023703}.

This versatility comes with inherent practical challenges. The success
probabilities for generating nG states lie between $10^{-6}$ and
$10^{-3}$ for typical parameters ($\epsilon\thicksim10^{-3}$, $g=0.01$),
a direct trade-off intrinsically linked to the large WVA necessary
for high-fidelity generation. For example, the heralded creation of
an enlarged odd SC state with fidelity $F_{-}>0.96$ from an input
with $\alpha=1.5$ occurs with a probability of $2\times10^{-4}$.
However, these probabilities fall within the practical regime enabled
by high-repetition-rate laser sources \citep{Ma_2015,XUE2024107048,CHEN2025111703},
which can transform low per-trial probabilities into high absolute
heralding rates of $10^{2}$ to $10^{4}$ states per second. This
positions the predicted performance of our protocol on par with the
demonstrated capabilities of other heralded nG state generation schemes,
such as SPAC \citep{2004} and cluster \citep{PhysRevLett.97.110501}
states, indicating that the achievable heralding rates would be on
a similar scale.

Nevertheless, the protocol places stringent demands on component performance.
Critically, the interferometer for the idler photon must exhibit high
phase stability to maintain the precise destructive interference condition
essential for the WVA effect; any instability introduces noise and
degrades the fidelity of the output state. Furthermore, photon loss
and non-ideal detector efficiency will introduce vacuum noise, reducing
output purity. Additionally, the requirement of a weak nonlinearity
($g\langle n_{b}\rangle\ll1$) constrains the usable interaction strength,
making the achievement of a practical heralding rate a significant
challenge.

Despite these limitations, our protocol establishes a valuable theoretical
and practical framework. Its feasibility is supported by the widespread
availability of high-repetition-rate laser sources, which are a standard
tool in advanced quantum optics for generating the required input
states and overcoming low probabilistic yields \citep{PhysRevLett.101.250501,2645,PhysRevA.110.023703}.
Looking forward, the convergence of this capability with continued
progress in low-loss integrated photonics \citep{PhysRevLett.133.083803},
high-efficiency superconducting detectors \citep{3803,Tao_2019,PhysRevApplied.19.064041},
and engineered giant nonlinearities \citep{2016} will directly enhance
its performance and scalability, making it a promising pathway for
advanced quantum state engineering.

In conclusion, we have shown that the synergistic use of WM and cross-Kerr
nonlinearity provides a powerful method for nG state generation. Remarkably,
this approach implicitly enacts nG operations, confirming the utility
of WM as a tool for quantum state engineering. The ability to generate
and amplify a continuous family of non-classical states from a unified
framework opens up new possibilities for applications in CV quantum
information processing.\vspace{0.3cm}

\textbf{METHODS }

\textbf{BS Transformation}

\begin{figure}
\includegraphics{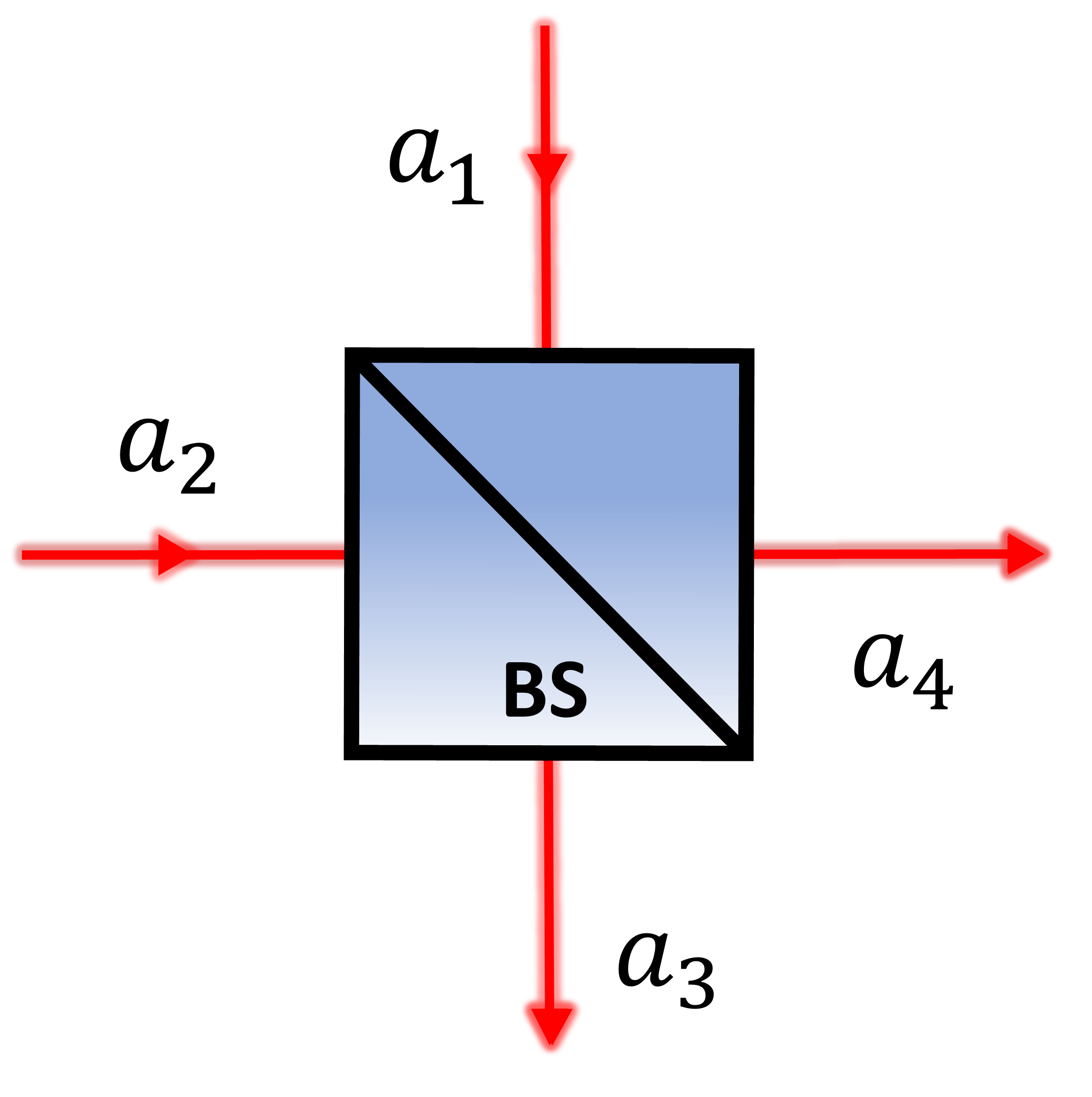}

\caption{\label{fig:11-1}\textbf{Schematic representation of a BS}. A BS scheme
with two input ports and two output ports. The annihilation operators
for the input modes are $a_{1}$ and $a_{2}$, and for the output
modes are $a_{3}$ and $a_{4}$, respectively.}

\end{figure}

The BS is a fundamental device in both classical and quantum optics,
functioning to split a beam of light into transmitted and reflected
components. The schematic of the BS used in our nG state generation
protocol is shown in Fig. \ref{fig:11-1}. It is well known that lossless
two-mode BS (with two input and output ports) in quantum optics is
described by the unitary matrix $U_{BS}$, which has the form 
\begin{equation}
\left(\begin{array}{c}
a_{3}\\
a_{4}
\end{array}\right)=U_{BS}(\theta,\varphi_{\tau},\phi_{\rho})\left(\begin{array}{c}
a_{1}\\
a_{2}
\end{array}\right),\label{eq:2}
\end{equation}
 where the explicit expression of $U_{BS}$ is given in Eq. (\ref{eq:16}),
and parameters $\theta,\varphi_{\tau}$ and $\phi_{\rho}$can be adjusted
for specific purposes. Here, the annihilation operators for the first
and second input modes are $a_{1}$ and $a_{2}$, respectively, and
for the output ports are $a_{3}$ and $a_{4}$. The action of the
BS transforms the states between the inner and outer paths of the
interferometer. For example, if we consider a single-photon in mode
$1$ and a vacuum state in mode $2$ of input ports, the BS transforming
the state $\vert1\rangle_{1}\vert0\rangle_{2}$ to 
\begin{align}
\vert\psi_{i}^{\prime}\rangle & =a_{1}^{\dag}\vert0\rangle_{1}\vert0\rangle_{2}\nonumber \\
 & =\left(e^{i\varphi_{\tau}}\cos\theta_{1}a_{3}^{\dag}-e^{-i\phi_{\rho}}\sin\theta_{1}a_{4}^{\dag}\right)\vert0\rangle_{3}\vert0\rangle_{4}\nonumber \\
 & =e^{i\varphi_{\tau}}\cos\theta_{1}\vert1\rangle_{3}\vert0\rangle_{4}-e^{-i\phi_{\rho}}\sin\theta_{1}\vert0\rangle_{3}\vert1\rangle_{4}.\label{eq:32-1}
\end{align}
The subindices of the states denote the corresponding paths in the
interferometer. Consequently, by setting $\varphi_{\tau}=0$, and
$\phi_{\rho}=\pi/2$,we obtain the preselection state $\vert\psi_{i}\rangle$
(see Eq. (\ref{eq:20-1})) used in our protocol. 

Furthermore, in the schematic of our protocol, a heralding event is
defined by a click at detector $D_{1}$ and no click at $D_{2}$ .
This corresponds to detecting a photon in one output port (e.g., arm
4) and no photon in the other (e.g., arm 3) of the second BS, which
is characterized by the parameters $\theta_{2}$, $\varphi_{\tau}^{\prime}$
and $\phi_{\rho}^{\prime}$. This is equivalent to selecting the following
path state inside the interferometer: 
\begin{align}
\vert\psi_{f}^{\prime}\rangle & =a_{D1}^{\dag}|0\rangle_{3}|0\rangle_{4}=a_{4}^{\dag}|0\rangle_{3}|0\rangle_{4}\nonumber \\
 & =(-e^{i\phi_{\rho}^{\prime}}\sin\theta_{2}a_{1}^{\dag}+e^{i\varphi_{\tau}^{\prime}}cos\theta_{2}a_{2}^{\dag})|0\rangle_{1}|0\rangle_{2}\nonumber \\
 & =e^{i\varphi_{\tau}^{\prime}}cos\theta_{2}|0\rangle_{1}|1\rangle_{2}-e^{i\phi_{\rho}^{\prime}}\sin\theta_{2}|1\rangle_{1}|0\rangle_{2}.\label{eq:33}
\end{align}
 With the specific parameters $\theta_{2}$=$\pi$/4, $\varphi_{\tau}^{\prime}=0,$
and $\phi_{\rho}^{\prime}=\pi/2$,the above state becomes the postselected
state $\vert\psi_{f}\rangle$ given in Eq. (\ref{eq:24}). 

\vspace{0.3cm}

\textbf{Numerical Simulation Details}

All theoretical results and state characterizations were obtained
through numerical simulations performed using the Quantum Toolbox
in Python (QuTiP) \citep{JOHANSSON20121760,JOHANSSON20131234}. The
simulation and analysis involved the following steps:

State Construction: The input states (coherent, squeezed vacuum, and
Schr\"{o}dinger cat states) were constructed using the relevant displacement
($D(\beta)$) and squeezing ( $S(r))$ operators within QuTiP. 

Protocol Implementation: The final state $\vert\Phi_{f}\rangle$ was
computed for each input state according to the derivation given in
Results sections.

Fidelity Calculation: The fidelity between the generated state $\rho_{out}=\vert\Phi_{f}\rangle\langle\Phi_{f}\vert$
and a target pure state $\vert\psi_{target}\rangle$ was computed
as $F=\vert\langle\psi_{target}\vert\Phi_{f}\rangle\vert^{2}$. 

Wigner Function Computation: The Wigner function of a state $\rho$
was calculated using the standard formula $\ensuremath{W(\alpha)=\frac{2}{\pi}Tr\left[D^{\dagger}(\alpha)\rho D(\alpha)(-1)^{b^{\dagger}b}\right]}.$
The Fock space dimension was set to 30 for simulations involving coherent,
squeezed vacuum and cat states to ensure the computed states were
well-contained within the Hilbert space and numerical truncation errors
were negligible.

\vspace{0.3cm}

\textbf{ACKNOWLEDGMENT}

This study was supported by the National Natural Science Foundation
of China (No. 12365005).

\textbf{AUTHOR CONTRIBUTIONS}

Xiao-Xi Yao derived and simulated the theoretical results; Yusuf Turek
wrote the main manuscript text and supervised the work . All authors
participated in the discussion of results and writing and revision
of the manuscript.

\textbf{COMPETING INTERESTS}

The authors declare no competing interests.

\textbf{DATA AVAILABILITY}

The source data that support the plots within this article and other
findings of this study are publicly available at https://github.com/YaoXiaoxi023/code-and-data. 

\textbf{ADDITIONAL INFORMATION}

\textbf{Correspondence} and requests for materials should be addressed
to Yusuf Turek. 

\textbf{REFERENCES}

\bibliographystyle{apsrev4-1}
\bibliography{Reference}
\vspace{0.3cm}

\end{document}